\documentclass[english,11pt]{article}
\pdfoutput=1
\usepackage{epsf,amsmath,amssymb,graphicx,dcolumn}
\usepackage[bottom]{footmisc}
\usepackage{scalefnt,subfigure,ulem}
\usepackage{cite,hyperref}
\usepackage{url}
\usepackage{todonotes}
\usepackage{color}
\usepackage{rotating}
\usepackage{multirow}
\definecolor{lightblue}{rgb}{.7,.8,1}
\title{\vspace*{-6em}
  \begin{flushright}
    \begin{minipage}{5cm}
    {\sf\small
      CERN-PH-TH-2014-170,\\[-5pt]
      WUB/14-07, ZU-TH 31/14,\\[-5pt]
      MCnet-14-19, LPN14-108\\[1em]
    }
    \end{minipage}
  \end{flushright}
\vspace*{2em}
Transverse momentum resummation for Higgs production via gluon fusion in the \mssm{}}
\author{Robert V. Harlander$^a$, Hendrik Mantler$^{b}$, Marius Wiesemann$^{c}$\\[2em]
 {\it $^a$Fachbereich C,
  Bergische Universit\"at Wuppertal,}\\[0em] {\it 42097 Wuppertal,
  Germany}\\[1em]
{\it $^b$TH Division, Physics Department, CERN}\\[0em]
{\it CH-1211 Geneva 23, Switzerland}\\[1em]
  {\it $^c$Physik-Institut, Universit\"at Z\"urich,}\\[0em]
 {\it 8057 Z\"urich, Switzerland}\\[1em]
{\small\tt robert.harlander@uni-wuppertal.de}\\[-.3em]
{\small\tt hmantler@cern.de}\\[-.3em]
{\small\tt mariusw@physik.uzh.ch}
}
\date{}

\def\bal#1\eal{\begin{align}#1\end{align}}
\newcommand{\bsquare}{pure-$b$}
\newcommand{\tsquare}{pure-$t$}
\newcommand{\tpsquarks}{no-$b$}
\newcommand{\interference}{int-$b$}

\newcommand{\abbrev}{\scalefont{.9}}

\newcommand{\mhiggs}{m_{\phi}}
\newcommand{\mh}{m_{h}}
\newcommand{\mA}{m_{A}}
\newcommand{\mH}{m_{H}}

\newcommand{\lhc}{{\abbrev LHC}}
\newcommand{\susy}{{\abbrev SUSY}}

\newcommand{\powheg}{{\abbrev POWHEG}}
\newcommand{\mcatnlo}{{\abbrev MC@NLO}}

\newcommand{\pt}{\ensuremath{p_T}}
\newcommand{\qt}{\ensuremath{p_T}}
\newcommand{\llog}{{\abbrev LL}}
\newcommand{\nll}{{\abbrev NLL}}
\newcommand{\nnll}{{\abbrev NNLL}}
\newcommand{\lo}{{\abbrev LO}}
\newcommand{\nlo}{{\abbrev NLO}}
\newcommand{\nnlo}{{\abbrev NNLO}}
\newcommand{\nlonll}{\nlo\plus\nll}

\newcommand{\sm}{{\abbrev SM}}
\newcommand{\qcd}{{\abbrev QCD}}
\newcommand{\thdm}{{\abbrev 2HDM}}
\newcommand{\mssm}{{\abbrev MSSM}}

\newcommand{\mb}{m_{\rm b}}
\newcommand{\mt}{m_{\rm t}}

\newcommand{\plus}{{\abbrev +}}
\newcommand{\citere}[1]{Ref.\cite{#1}}
\newcommand{\citeres}[1]{Refs.\cite{#1}}
\newcommand{\eqn}[1]{Eq.\,(\ref{#1})}
\newcommand{\neqn}[1]{Eqs.\,(\ref{#1})}
\newcommand{\fig}[1]{Fig.\,\ref{#1}}
\newcommand{\figs}[1]{Figs.\,\ref{#1}}
\newcommand{\tab}[1]{Tab.\,\ref{#1}}
\newcommand{\sct}[1]{Section~\ref{#1}}

\newcommand{\dd}{{\rm d}}

\newcommand{\ccbar}{c\bar c}

\newcommand{\als}{\ensuremath{\alpha_s}}

\newcommand{\api}{\frac{\alpha_s}{\pi}}
\newcommand{\muF}{\mu_{\rm F}}
\newcommand{\muR}{\mu_{\rm R}}
\newcommand{\Qres}{Q}

\newcommand{\pdf}{{\abbrev PDF}}

\newcommand{\bld}[1]{\boldmath{$#1$}}

\newcommand{\tauphobic}{$\tau$-phobic}
\newcommand{\lightstop}{light-stop}
\newcommand{\lightstau}{light-stau}
\newcommand{\mhmodp}{$m_h^{\text{mod}+}$}
\newcommand{\mhmodm}{$m_h^{\text{mod}-}$}
\newcommand{\mhmax}{$m_h^{\text{max}}$}
\newcommand{\scenario}{\ensuremath{\mathcal S}}

\newcounter{notecount}
\begin{document}
\maketitle

\vspace*{1cm}


\begin{abstract}
\noindent
The resummed transverse momentum distribution of supersymmetric Higgs
bosons produced through gluon fusion at \nlo{}\plus\nll{} is presented,
including the exact quark and squark mass dependences. Considering
various \mssm{} scenarios, we compare our results to previous ones
within the \powheg{} approach. We analyze the impact of the bottom loop
which becomes the dominant contribution to the gluon fusion cross
section for a wide range of the parameter space for the pseudo-scalar
and heavy Higgs.
\end{abstract}
\vfill


\thispagestyle{empty}

\clearpage


\section{Introduction}
\label{sec:intro}
After the observation of a Higgs boson of mass $125$\,GeV
\cite{Aad:2012tfa,Chatrchyan:2012ufa}, the measurement of its properties
has become one of the central targets of the \lhc{}. From the
theoretical side, precise predictions for the production and decay rates
of such a particle in various models are crucial to pin down its
nature. An enormous effort has already gone into precision calculations
of the total cross section as well as kinematical distributions (see
\citeres{Dittmaier:2011ti,Dittmaier:2012vm,Heinemeyer:2013tqa} for an
overview).

In the Standard Model (\sm{}), the Higgs is predominantly produced via
gluon fusion, where the Higgs-gluon coupling is mediated by a quark
loop. Its cross section is about an order of magnitude higher than the
sum of all other processes, which retain their importance through their
additional final state particles and/or their specific kinematics. The
gluon fusion process has been studied in great detail over the recent
years, leading to a significant decrease of the related theoretical
uncertainties. In particular, the use of an effective theory approach
for the calculation of higher order corrections allows, loosely
speaking, to determine the cross section one perturbative order higher
than in the full theory. Within this approach, also known as the
heavy-top limit, the top quark is assumed to be infinitely heavy. The
total cross section in this approximation is known up to
next-to-next-to-leading order (\nnlo{})
\cite{Harlander:2002wh,Anastasiou:2002yz,Ravindran:2003um} and even
parts of the next-to-\nnlo{} are known
\cite{Ball:2013bra,Buehler:2013fha,Boughezal:2013uia,Anastasiou:2014vaa}.
Electro-weak corrections and further effects beyond \nnlo{} have been
evaluated in
\citeres{Catani:2003zt,Idilbi:2005ni,Ravindran:2006cg,Ahrens:2008nc,
  Djouadi:1994ge,Degrassi:2004mx,Aglietti:2004nj,Actis:2008ug,
  Anastasiou:2008tj} for example.  The uncertainty induced by the
heavy-top limit has been shown to be below 1\% for the total rate at
\nnlo{}\cite{Marzani:2008az,Harlander:2009mq,Harlander:2009my,
Pak:2009dg,Pak:2011hs}.

While the effects of the four lightest quarks as mediators of the
gluon-Higgs coupling is negligible ($\lesssim 1$\%) and therefore
usually omitted, the bottom quark contributes at the $5-10$\% level to
the total cross section at next-to-leading order (\nlo{})
\cite{Spira:1995rr,Harlander:2003xy}.  Due to the smallness of the
bottom-quark mass, one cannot apply the same approximation as for the
top-quark contributions to evaluate radiative corrections for the bottom
loop, but typically includes the full quark mass dependence in the
calculation.

Kinematical distributions of the Higgs boson provide an important handle
on the determination of Higgs properties (see, e.g.,
\citeres{Harlander:2013oja,Grojean:2013nya,Azatov:2013xha}). One of the
most important differential observables in this respect is the
transverse momentum ($\pt$) distribution of the Higgs. Once sufficient
statistics have been collected at the \lhc{}, the comparison of the
experimental result for this spectrum to its theoretical prediction in
various models should allow for further restrictions of the allowed
parameter space of these models.

In the \sm{}, the \nnlo{} transverse momentum distribution of the Higgs
produced via gluon fusion at $\pt{}>0$ has already been known for some
time in the heavy-top limit
\cite{deFlorian:1999zd,Ravindran:2002dc,Glosser:2002gm}.\footnote{Throughout
  this paper, we consistently associate the $2\to 1$ process
  $gg\rightarrow H$ with the leading-order (\lo{}) $\pt{}$ distribution,
  although it only contributes at $\pt{}=0$.}  Sub-leading top-mass
effects have been considered in
\citeres{Harlander:2012hf,Neumann:2014nha}. Furthermore, also the
fully-differential cross section has been determined up to \nnlo{}
\cite{Anastasiou:2004xq,Catani:2007vq,Catani:2008me}. However, it is
well known that those perturbative calculations break down for small
transverse momenta due to the occurrence of logarithmically enhanced
terms in $\pt$. Only a resummation of these terms to all orders in
$\als$ provides a proper theoretical prediction at small values of
$\pt$.

Transverse momentum resummation at leading logarithmic (\llog{}) and
next-to-\llog{} (\nll{}) accuracy for the gluon fusion process in the heavy-top 
approximation has already been performed a long time ago
\cite{Catani:1988vd,Yuan:1991we,Kauffman:1991cx}. \citere{Bozzi:2005wk}
introduced a matching procedure to consistently combine the resummed
distribution and the fixed order cross section valid at large $\pt$. Its
application to the $\pt$ spectrum of the Higgs at \nnlo{}\plus{}\nnll{}
was implemented using the effective theory approach
 into the publicly available program {\tt HqT}
\cite{Bozzi:2003jy,Bozzi:2005wk,deFlorian:2011xf}.\footnote{A Monte
  Carlo approach, based on the same resummation formalism, to add
  resummation effects to the differential \nnlo{} cross section with respect 
  to the Higgs and its decay products was
  implemented into the program {\tt HRes} \cite{deFlorian:2012mx,Grazzini:2013mca}.} 

Finite top- and bottom-mass effects on the resummed $\pt$ spectrum have
been considered in the \powheg{}\cite{Nason:2004rx}
approach\cite{Bagnaschi:2011tu} and by analytic resummation through
\nlo{}\plus\nll{}\cite{Mantler:2012bj,Grazzini:2013mca,Banfi:2013eda}.
The small bottom-quark mass $\mb$ introduces an additional uncertainty
because terms $\ln(\mb/\pt{})$ appear at the amplitude
level\footnote{However, in the limit $\pt\to 0$, these terms
  $\sim\ln(\mb/\pt{})$ vanish, and therefore collinear and soft
  factorization is preserved.}, which are potentially large and could
spoil the collinear and soft approximation already at $\pt{}\gtrsim \mb$
\cite{Grazzini:2013mca}. The small bottom Yukawa coupling in the \sm{}
prevents this uncertainty from becoming too severe though.

Supersymmetric extensions are among the most popular theories beyond the
\sm{}. The minimal supersymmetric \sm{} (\mssm{}) contains two Higgs
doublets, which lead to five physical Higgs bosons, three of which are
neutral and two charged. The production of the neutral \mssm{} Higgs
bosons is typically dominated by either of two processes, gluon fusion
or bottom-quark annihilation. For the theoretical status of the latter
process, we refer the reader to \citere{Harlander:2014hya}, where the
resummed $\pt$ distribution through \nnlo{}\plus{}\nnll{} was obtained,
and the references therein.\footnote{The \nnlo{} \pt{} distribution for
  bottom-quark annihilation is already known for a while
  \cite{Harlander:2010cz,Ozeren:2010qp,Harlander:2011fx,Buehler:2012cu}.}

In this paper, we focus on the gluon fusion process in the \mssm{}, but
our calculation will be applicable also in a general 2-Higgs doublet
model (\thdm). The
total Higgs production cross section in gluon fusion in the \mssm{} has
been calculated up to \nlo{} within the
\mssm{}\cite{Harlander:2003bb,Harlander:2004tp,Harlander:2005if,Muhlleitner:2006wx,Degrassi:2008zj,Anastasiou:2008rm,Harlander:2010wr,Degrassi:2010eu,Muhlleitner:2010nm,Degrassi:2011vq,Bagnaschi:2011tu,Degrassi:2012vt}. The currently most accurate
total cross section in the \mssm{} can be obtained with the publicly
available program {\tt SusHi} \cite{Harlander:2012pb}.  It is obvious
that our approach is also directly applicable to all neutral Higgs
bosons of the \thdm{}.\footnote{Concerning the
  total cross section in \thdm{}s, see \citere{Harlander:2013qxa}.}

Our goal is the analytically resummed $\pt$ spectrum of all three
neutral \mssm{} Higgs bosons produced through gluon fusion at
\nlo{}\plus{}\nll{}. Since the bottom Yukawa coupling can be
significantly enhanced with respect to the \sm{}, the issue of a proper
treatment of bottom-quark induced effects on the cross section becomes
more important. We propose a pragmatic way to separately set the
resummation scale of these terms and to derive an estimate of the
residual uncertainty.

We compare our results to the ones of a similar earlier study
\cite{Bagnaschi:2011tu}, which calculated the transverse momentum
spectrum within the \powheg{} approach \cite{Alioli:2008gx} in
combination with a parton shower.

The paper is organized as follows: In \sct{sec:resum}, we briefly review
the formalism for the resummation of contributions at small transverse
momenta in the gluon fusion process and discuss the required theoretical
quantities. Our procedure for choosing the resummation scale is
described in \sct{sec:setres}. \sct{sec:input} lists the input
parameters and defines a set of \mssm{} parameter points which we use
for our analysis. It also describes the way we determine the theoretical
uncertainties.  Numerical results are presented in \sct{sec:results},
where we analyze the $\pt$ spectra for all three neutral \mssm{} Higgs
bosons in specific scenarios and the impact of the relative
contributions ordered by the respective Yukawa couplings that enter the
cross section.  \sct{sec:conclude} contains our conclusions.


\section{Transverse momentum resummation}
\label{sec:resum}


\subsection{Resummation and Matching}
\label{sec:matching}
Consider the transverse momentum ($\pt{}$) distribution of a
color-neutral heavy particle of mass $M$ produced via a $2\rightarrow 1$
process in \qcd{}. For $\pt{}\gtrsim M$, a fixed-order expansion of the
cross section in the strong coupling $\als$ can be applied. In the limit
$\pt{}\rightarrow 0$, however, large logarithms $\ln(\pt{}/M)$ appear at
fixed order, which spoil the validity of the perturbative expansion. A
proper prediction of the distribution at $\pt\ll M$ can be obtained by
resumming these logarithms to all orders in $\als$. Following
\citere{Bozzi:2005wk}, we split the $\pt{}$-dependent cross section as
\begin{align}
\label{eq:match}
\frac{\dd\sigma}{\dd\pt^2}=\frac{\dd\sigma^{\text{(res)}}}{\dd\pt^2} +
\frac{\dd\sigma^{\text{(fin)}}}{\dd\pt^2}\,,
\end{align}
where the resummed logarithmic contributions in $\pt$ are contained in
the first term on the r.h.s., while the second term remains finite as
$\pt\to 0$.  
Working at finite orders, the cross section can be cast into the
following form:
\begin{align}
\label{eq:match-order}
\left[\frac{\dd\sigma}{\dd\pt^2}\right]_{\text{f.o.}+\text{l.a.}}=
\left[\frac{\dd\sigma^{\text{(res)}}}{\dd\pt^2}\right]_{\text{l.a.}}
+\left[\frac{\dd\sigma}{\dd\pt^2}\right]_{\text{f.o.}}
-\left[\frac{\dd\sigma^{\text{(res)}}}{\dd\pt^2}\right]_{\text{f.o.}}\,.
\end{align}
where ``f.o.''(=fixed order) denotes the perturbative, and ``l.a.'' the
logarithmic accuracy (to be defined below) under consideration. The
imposed matching condition
\begin{equation}
\begin{split}
\left[ \left[\frac{\dd\sigma^{\text{(res)}}}{\dd\pt^2}\right]_{\text{l.a.}}
  \right]_{\text{f.o}} =
\left[\frac{\dd\sigma^{\text{(res)}}}{\dd\pt^2}\right]_{\text{f.o.}}\,,
\label{eq:matchcond}
\end{split}
\end{equation}
defines the logarithmic accuracy needed at a specific perturbative
order, and vice versa. In \eqn{eq:match-order}, all terms
$\sim\delta(\pt{})$ are contained in the first term of the r.h.s.; in
practical calculations, one can therefore disregard such terms in the
second and third term since they will cancel among each other.

The matching procedure as proposed in \citere{Bozzi:2005wk} induces a
{\it unitarity constraint} on the matched cross section which implies
that the integral over $\pt^2$ reproduces the total cross section
$\sigma_\text{tot}$ at fixed order:
\begin{align}
\int \dd\pt^2 \left[\frac{\dd\sigma}{\dd\pt^2}\right]_{\text{f.o.+l.a.}}
&\equiv\left[\sigma_\text{tot}\right]_{\text{f.o.}}\,.
\label{eq:unitarity}
\end{align}
In the next section, we will address the evaluation of
$\dd\sigma^\text{(res)}/\dd \pt^2$.

\subsection{The resummed cross section}
\label{sec:elements}
The resummation of large logarithmic contributions is performed in the
impact parameter or $b$ space, given by the Fourier transform
w.r.t.\ the transverse momentum:\footnote{Throughout this paper,
  parameters that are not crucial for the discussion will be suppressed
  in function arguments. Note that we refrain from including the spin
  correlation functions $G$ introduced in \citere{Catani:2010pd} here
  and in what follows, since they are not required at the order
  considered in this paper.}\cite{Collins:1984kg,Catani:2000vq}
\begin{equation}
\begin{split}
\label{eq:res}
\frac{\dd \sigma^{F,\text{(res)}}}{\dd \qt^2}&= \tau \int_0^{\infty} \dd
b\, \frac{b}{2} \,J_{0}(b\,\qt) \,
\sum_{c\in\{g,q,\bar{q}\}}\hat\sigma^{F,(0)}_{\ccbar}\, H_c^F(\als) \,
S_c(M,b)\\ &\, \times \sum_{i,j\in\{g,q,\bar{q}\}}
\,\left[C_{ci}(\als(b_0/b))\otimes C_{\bar{c}j}(\als(b_0/b))\otimes
f_i\left(b_0/b\right)\otimes f_j \left(b_0/b\right)\right](\tau)\,,
\end{split}
\end{equation}
with $q\in\{u,d,s,c,b\}$, a numerical
constant\footnote{$\gamma_E=-\Gamma'(1)$ is the Euler constant.}
$b_0=2\exp(-\gamma_E)=1.12292\ldots$, and the Bessel function of the
first kind $J_0(x)$ with $J_0(0)=1$. Here and in what follows, the
superscript $F$ is attached to process specific quantities in order to
distinguish them from universal ones.  The symbol $\otimes$ indicates
the convolution in the following sense:
\begin{equation}
\begin{split}
&\left[C_{ci}(\als(b_0/b))\otimes f_j(b_0/b)\right](z_3) \equiv\\&\quad\equiv
\int_0^1\dd z_1\int_0^1\dd z_2\,\delta(z_3-z_1z_2)\,C_{ci}(\als(b_0/b),z_1)\,
f_j(z_2,b_0/b),
\end{split}
\end{equation}
where $f_j(x,q)$ denotes the parton $i$ with momentum fraction $x$
of the proton, and evaluated at momentum transfer $q$.

The central element of the resummation formula is the so-called Sudakov
form factor
\begin{equation}
\begin{split}
\label{eq:sudakov}
S_c(M,b)&=\exp\left\{L\,g_c^{(1)}(\als\,L)+g_c^{(2)}(\als\,L)
+\sum\limits_{l=3}^\infty\left(\api\right)^{l-2}\,
g_c^{(l)}(\als\,L)\right\},
\end{split}
\end{equation}
which resums logarithms of the form $L=\ln(b^2M^2/b_0^2)$, while
$\als\,L$ is treated as being of order unity. The order of the expansion
in the exponent then defines the {\it logarithmic accuracy}. At
  leading logarithmic level, only $g_c^{(1)}$ has to be taken
into account, at \nll{} also $g_c^{(2)}$ and so
forth. We give their functional expressions up to the required order in
this paper (i.e. $g_c^{(1)}$ and $g_c^{(2)}$) in \ref{app:gs}.

Clearly, there is a certain amount of freedom in the separation between 
the ``hard'' and the ``soft'' region which can be parametrized by the
so-called {\it resummation scale} $Q$. Unless indicated otherwise, we have set
$Q\equiv M$ throughout this section; the generalization of the formulas to
$Q\neq M$ and consequently $L=\ln(Q^2b^2/b_0^2)$ can be found in
\citere{Bozzi:2005wk}. In fact, the choice of the resummation scale for
the gluon fusion process will be one of the central issues of this paper
and will be discussed in more detail in \sct{sec:setres}.

The Born factor $\hat\sigma_{\ccbar}^{F,(0)}$ in \eqn{eq:res} is given by the parton
level cross section at \lo{}.  In general, the sum over $c$ accounts for
all \lo{} subprocesses that can produce the considered colorless
particle. In the gluon fusion process though, only $c=g$ is relevant. An
explicit analytical expression for $\hat\sigma_{gg}^{F,(0)}$ for this
process can be found in Eq.\,(21) of \citere{Harlander:2012pb}, for
example.\footnote{In the notation of \citere{Harlander:2012pb}, it is
  $\sigma_{\ccbar}^{F,(0)} \equiv \sigma_0^\phi$.}

The {\it resummation coefficients} in \eqn{eq:res} can be expanded perturbatively:
\begin{equation}
\begin{split}
C_{ci}(\als,z) &= \delta_{ci}\delta(1-z) + \sum_{n=1}^{\infty}
\left(\api \right)^n C_{ci}^{(n)}(z)\,,\qquad H_c^F(\als) =
1+\sum_{n=1}^{\infty} \left(\api \right)^n H_c^{F,(n)}\,.
\label{eq:rescoef}
\end{split}
\end{equation}
\nll{} accuracy requires the knowledge of these coefficients to first
order in $\als$. Evidently, $\delta(1-z)$ terms in $C_{cj}^{(n)}$ can be
shifted to $H_c^{(n),F}$ for $n\ge1$, and vice versa. A ￼particular
choice of these terms in $C_{cj}^{(n)}$ (equivalently, a particular
choice of $H_c^F$ for one process) defines what is called a {\it
  resummation scheme}\cite{Bozzi:2005wk}. Within a particular
resummation scheme, the coefficients $C_{cj}$ can be considered
universal, while $H_c^F$ is process dependent. The entire process
dependence in \eqn{eq:res} within a given resummation scheme is thus
contained in $H_c^F$ and $\sigma_{c\bar{c}}^{F,(0)}$.  We give the
resummation coefficients in the $gg\rightarrow\phi$ scheme
($\phi\in\{h,H,A\}$), which is defined by setting
\begin{equation}
\begin{split}
H_g^F(\als)\equiv 1
\end{split}
\end{equation}
for
this process.

The $C$-coefficients in the  $gg\rightarrow\phi$ scheme are \cite{deFlorian:2001zd}
\begin{equation}
\begin{split}
\label{eq:rescoeff2}
C_{gq}^{(1)}(z) &=C_{g\bar{q}}^{(1)}(z)=\frac{C_F}{2}\,
z\,,\qquad C_{gg}^{(1)}(z) = \delta(1-z)\left(\frac{C_A}{2}
\zeta_2+\frac{\mathcal{A}^{\phi,\text{virt}}_g}4\right)\,,
\end{split}
\end{equation}
where again $q\in\{u,d,s,c,b\}$ and $\zeta_2\equiv\zeta(2)=\pi^2/6$, with
Riemann's $\zeta$ function. $\mathcal{A}^{\phi,\text{virt}}_g$ denotes the finite part of the virtual corrections as defined in Eq. (38) of \citere{deFlorian:2001zd}, i.e.
\begin{equation}
\begin{split}
\mathcal{A}^{\phi,\text{virt}}_g = \left. 2\cdot C^{\phi}\right |_{\muR=\muF}
\end{split}
\end{equation}
with $C^{\phi}$ from Eq. (27) in \citere{Harlander:2012pb}. 

It can be shown that the unitarity constraint of \eqn{eq:unitarity} can
be imposed by replacing
\begin{equation}
\begin{split}
L\to \widetilde L \equiv \ln\left(\frac{Q^2b^2}{b_0^2}+1\right)\,,
\label{eq:Ldef}
\end{split}
\end{equation}
in \eqn{eq:sudakov}.  In addition, this replacement reduces unjustified
resummation effects at high transverse momenta, since $\widetilde L$
vanishes in the limit $b\rightarrow 0$ (i.e. $\pt{}\rightarrow \infty$),
while the large-$b$ limit (small $\pt{}$) is preserved.

With the replacement in \eqn{eq:Ldef} the resummed cross section $\dd\sigma^\text{(res)}/\dd\pt^2$ becomes
explicitely $\Qres$ dependent; however, this dependence formally cancels between
$[\dd\sigma^\text{(res)}/\dd\pt^2]_{\text{l.a.}}$ and
$[\dd\sigma^\text{(res)}/\dd\pt^2]_{\text{f.o.}}$ in \eqn{eq:match-order}. Any
residual dependence of the final result is beyond the specific
logarithmic order under consideration.  The variation
of the cross section with $\Qres$ will be used to estimate the
uncertainty due to missing terms of higher logarithmic accuracy.


\subsection{Components to the matched-resummed cross section}
\label{sec:outline}
The goal of this paper is to determine $\pt{}$ spectra of neutral
\mssm{} Higgs bosons produced via gluon fusion by matching the
\nlo{} result to the resummed \nll{} approximation.

\begin{figure}[h]
  \begin{center}
    \begin{tabular}{ccccc}
      \mbox{\includegraphics[height=.12\textheight]{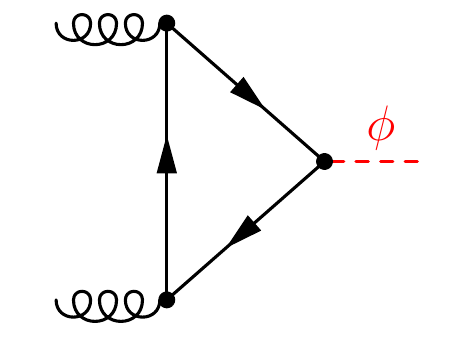}} & & \mbox{\includegraphics[height=.12\textheight]{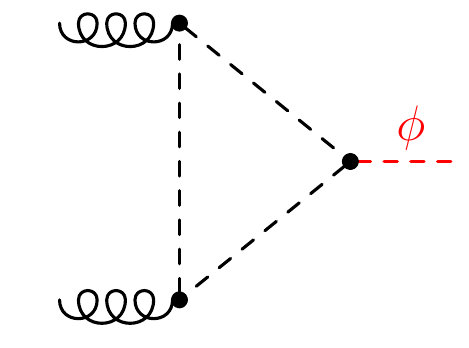}} & &
      \mbox{\includegraphics[height=.12\textheight]{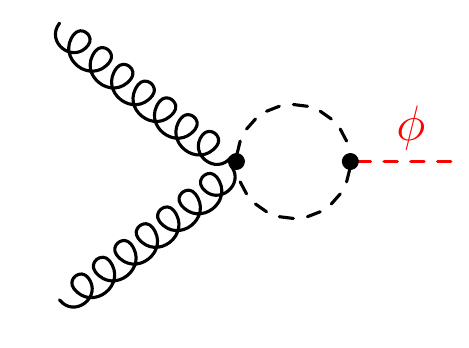}}
            \\
      (a) & & (b) & & (c)\\
      \mbox{\includegraphics[height=.12\textheight]{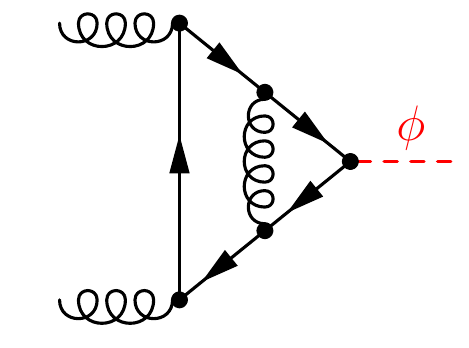}} & & \mbox{\includegraphics[height=.12\textheight]{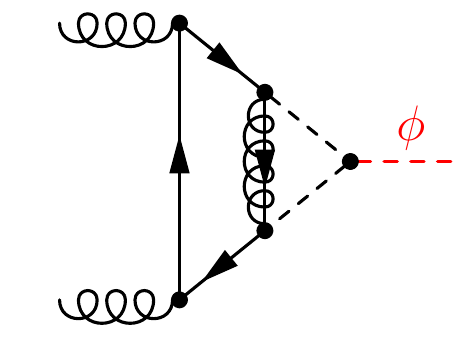}} & &
      \mbox{\includegraphics[height=.12\textheight]{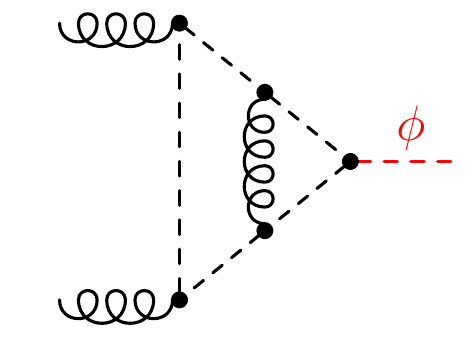}}
            \\
      (d) & & (e) & & (f)\\
      \mbox{\includegraphics[height=.12\textheight]{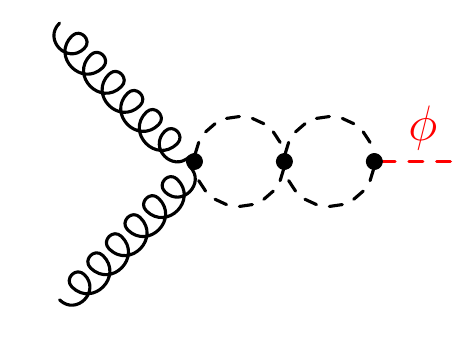}} & & \mbox{\includegraphics[height=.085\textheight]{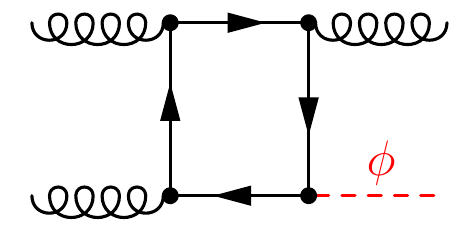}} & &
      \mbox{\includegraphics[height=.11\textheight]{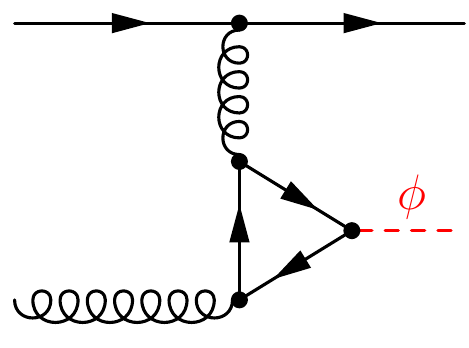}}
            \\
      (g) & & (h) & & (i)
    \end{tabular}
    \parbox{.9\textwidth}{%
      \caption[]{\label{fig:diag} A sample of Feynman diagrams for
        $gg\rightarrow \phi$ contributing to the \nlo{} cross section;
        (a-c) \lo{}, (d-g) virtual and (h-i) real corrections. The
        graphical notation for the lines is: solid straight
        $\widehat{=}$ quark; spiraled $\widehat{=}$ gluon; dashed
        $\widehat{=}$ scalar (squark or Higgs); spiraled with line
        $\widehat{=}$ gluino.  } }
  \end{center}
\end{figure}

The relevant \nlo{} matrix elements are taken from
\citere{Harlander:2012pb}, which include the \sm{}-like contributions as
well as sbottom, stop and gluino effects (see \fig{fig:diag} for some
sample Feynman diagrams). The \lo{} diagrams,
e.g. \fig{fig:diag}\,(a)-(c), determine the Born factor
$\hat\sigma_{gg}^{\phi,(0)}$. The \nlo{} \pt{} distribution
$[\dd\sigma/\dd\pt^2]_\text{f.o.=\nlo{}}$ in \eqn{eq:match-order} at
$\pt{}>0$\footnote{Note that $\delta(\pt)$ terms can be disregarded, see
  \sct{sec:elements}.} is governed by the real emission diagrams like
the ones shown in \fig{fig:diag}\,(h) and (i) (and similar ones with
quark loops replaced by squark loops). Finally, the virtual diagrams,
e.g. \fig{fig:diag}\,(d)-(g), enter $C^{(1)}_{gg}$, as can be seen from
\eqn{eq:rescoeff2}. These contributions allow to calculate
$[\dd\sigma^\text{(res)}/\dd\pt^2]_\text{l.a.=\nll}$.

The expansion of $\dd\sigma^{\text{(res)}}/\dd\pt^2$ with respect to \als{}
determines the logarithmic terms at \nlo{} in \eqn{eq:match-order}. The
explicit expression can be found in Eq.\,(72) of \citere{Bozzi:2005wk},
with the corresponding coefficients in Eq.\,(63) and (64) of that paper.

The resummed expression $[\dd\sigma^{\text{(res)}}/\dd\pt^2]_\text{\nll{}}$
has been calculated with a modified version of the program {\tt
  HqT}~\cite{Bozzi:2003jy,Bozzi:2005wk,deFlorian:2011xf}, which
determines the \nnlo{}\plus{}\nnll{} $\pt$ distribution for the
gluon fusion process using the approximation of an infinitely heavy top
quark. We modified it for our purposes and implemented the resummation
coefficients of \neqn{eq:rescoeff2} to include the \mssm{} effects.

\section{Choosing the resummation scale}
\label{sec:setres}

While the matched cross section is formally independent of the
resummation scale $\Qres$, the actual numerical result can be quite
sensitive to its particular choice due to the truncation at finite
logarithmic order. It is therefore vital to
determine an ``optimal'' choice for this scale. 
The resummation scale $\Qres$ can be viewed as a scale up
to which resummation is extended.
The soft and collinear approximation can be trusted only up to a finite value of
$\pt$ which is determined by a characteristic external scale of the problem. 
Consequently, there is a maximum value of
$\pt$ above which resummation is not valid and therefore, $\Qres$ should 
not be chosen beyond this value. Due to the constraint of
\eqn{eq:unitarity}, a too large value of $\Qres$ not only spoils the
prediction for $\pt<\Qres$, but also affects the large-$\pt$
region\footnote{By ``large-\pt{}'', we mean transverse momenta ``of the
  order of the characteristic scale'' and beyond.}. The distribution can
thus deviate significantly from the fixed-order prediction even in
regions where the latter should provide a good approximation.

For the top-quark induced gluon-Higgs coupling, the characteristic scale
is of the order of the Higgs mass ($\mhiggs$). Consequently, a reasonable range for
phenomenological studies is $\Qres\in[\mhiggs/4,\mhiggs]$, for
example. The need for precise predictions requires one to take into
account also bottom-quark induced effects to the gluon-Higgs coupling
though\cite{Mantler:2012bj,Bagnaschi:2011tu}.  It has been shown
\cite{Grazzini:2013mca} that terms $\sim\ln(\mb/\pt)$ appear in the
amplitude, which are potentially large and thus could spoil the soft and
collinear approximation for $\pt>\mb$. This has been used as an argument
to choose $\Qres=\mb$ \cite{Grazzini:2013mca}, i.e. to effectively turn
off Sudakov resummation at $\pt\gtrsim \mb$, even though their actual
impact has not been studied quantitatively in this context. However,
these logarithms are not of Sudakov type as they vanish when
$\pt\rightarrow 0$. In fact, they are closely related to logarithms
$\ln(\mb/\mh)$ which induce an uncertainty already at the level of the
total cross section\cite{Spira:1995rr}.  For a related quantity, namely the
cross section with a veto on jets with
$\pt^\text{jet}>p_{T,\text{veto}}^\text{jet}$, it was argued
\cite{Banfi:2013eda} that the impact of the analogous terms
$\sim\ln(\mb/p_{T,\text{veto}}^\text{jet})$ remains moderate, and that
one can treat these logarithms as a ``finite remainder'' together with
all other finite terms (power corrections in
$p_{T,\text{veto}}^\text{jet}$). A similar argument should apply also
to the Higgs' $\pt$ spectrum.

In this section, we will formulate a pragmatic though quite general way
to set the resummation scale. Rather than providing an {\it ex-ante}
value of $\Qres$ for a particular process, our method is trial-and-error
based and relies on simple expectations on the properties of the matched
$\pt$ distribution. Roughly speaking, we determine the value of $\Qres$
as large as possible while requiring that the large-$\pt$ behavior of
the matched distribution stays reasonably close to the fixed-order
prediction.

More precisely, for Higgs masses up to $\mhiggs=300$\,GeV, we determine
$\Qres^\text{max}$ as the maximum value of $\Qres$ for which the
resummed $\pt$-distribution stays within the interval
\mbox{$\left[0,\!2\right]\!\cdot\!  [\dd\sigma\!/\dd\pt^2]_\text{f.o.}$}
for $p_T\in[\mhiggs,\pt^\text{max}]$. The restriction to the latter
interval is needed because, on the one hand, resummation effects are
expected to be large for smaller values of $\pt$; on the other hand, the
numerical accuracy of our implementation of the resummation formula
becomes unreliable above certain values of $\pt$. The specific value of
$\pt^\text{max}$ needs to be chosen case by case. For
$\mhiggs=125.6$\,GeV, it is $\pt^\text{max}\approx 400$\,GeV, for
$\mhiggs=300$\,GeV, we use $\pt^\text{max}\approx 650$\,GeV.

\begin{figure}
  \begin{center}
\hspace*{-0.38cm}
    \begin{tabular}{cc}
    \hspace{-1.4cm}
    \includegraphics[angle=0,width=0.64\textwidth]{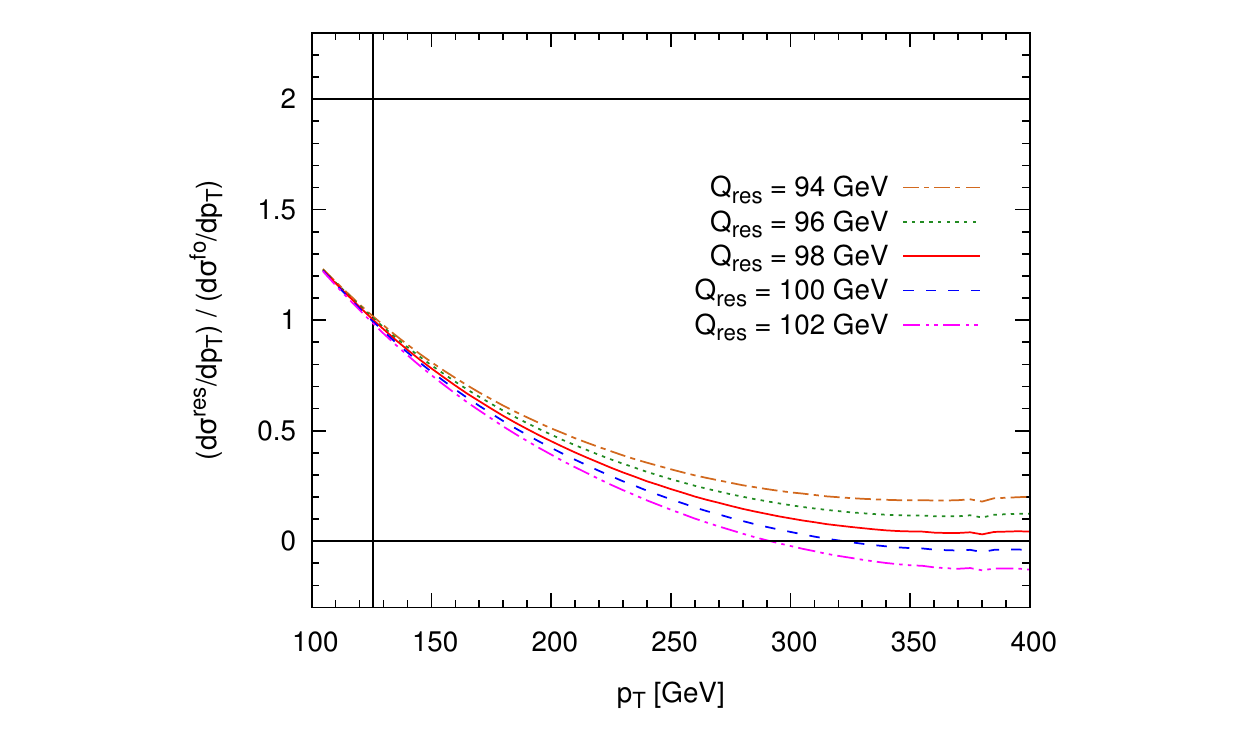}
    &\hspace{-2.73cm}
    \includegraphics[angle=0,width=0.64\textwidth]{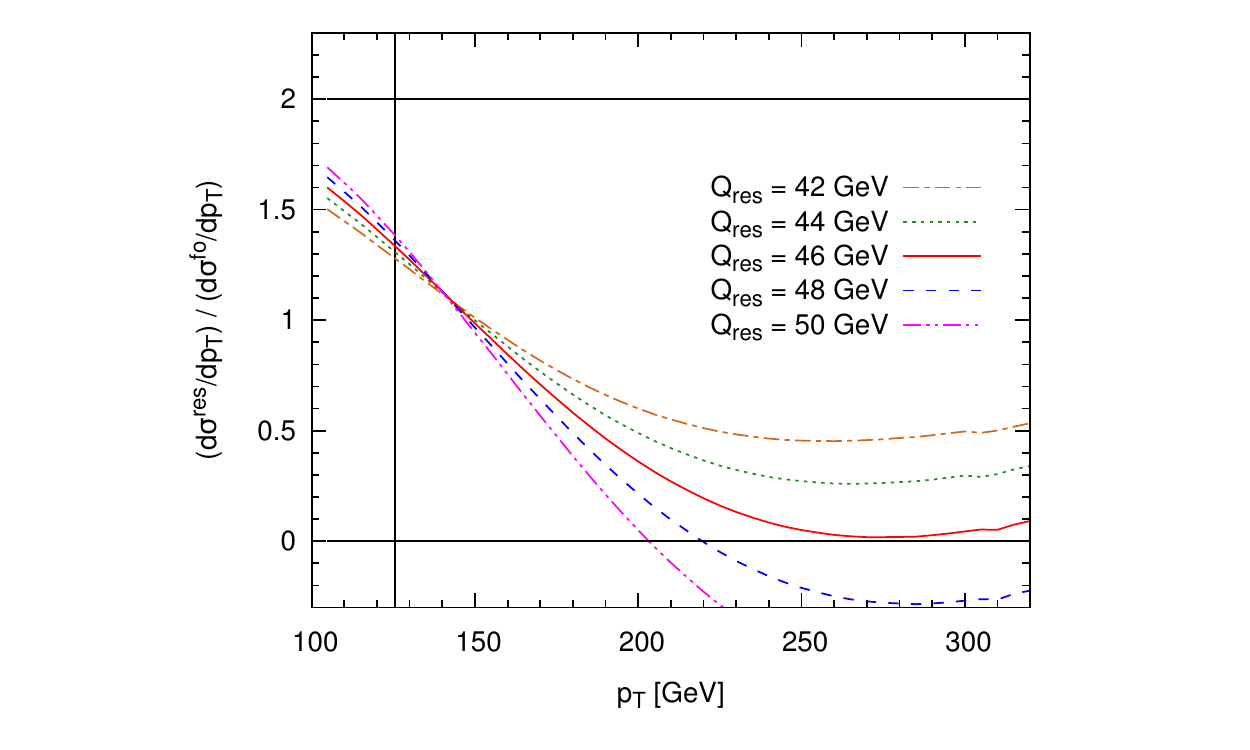}\\[-0.3cm]
    \hspace{-0.7cm} (a) & \hspace{-2cm} (b)\\[0.5cm]
    \multicolumn{2}{c}{%
    \hspace{-1.4cm}
    \includegraphics[angle=0,width=0.64\textwidth]{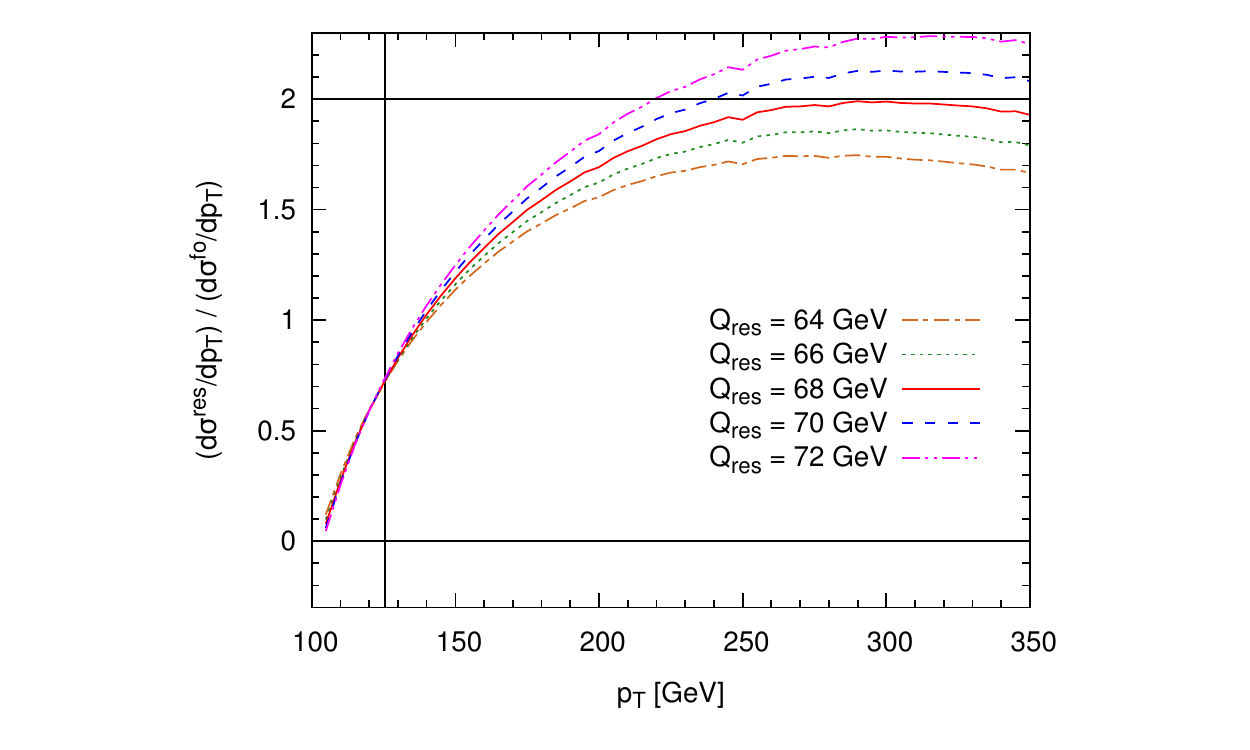}}\\[-0.3cm]
    \multicolumn{2}{c}{\hspace{-0.67cm} (c)}
    \end{tabular}
  \parbox{.9\textwidth}{%
  \caption{ \label{fig:qres} Large $\pt$-behavior of the cross section
    for a {\abbrev CP}-even Higgs boson with $\mh=125.6$\,GeV. The
    different lines correspond to various choices of the resummation
    scale. (a)\,Pure top quark, (b)\,pure bottom quark, and (c)
    top-bottom interference contribution. The vertical line marks the
    value of the Higgs mass.}}
  \end{center}
\end{figure}

\begin{figure}
  \begin{center}
\hspace*{-0.38cm}
    \begin{tabular}{cc}
    \hspace{-1.4cm}
    \includegraphics[angle=0,width=0.64\textwidth]{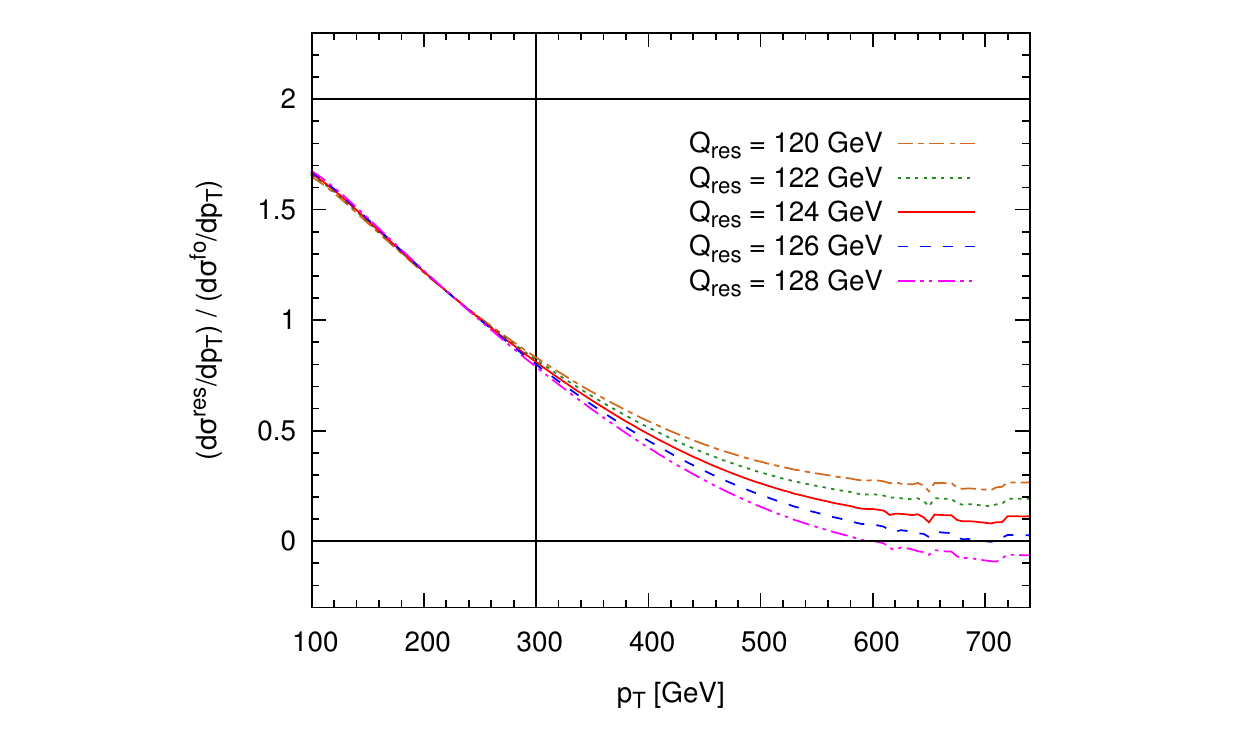}
    &\hspace{-2.73cm}
    \includegraphics[angle=0,width=0.64\textwidth]{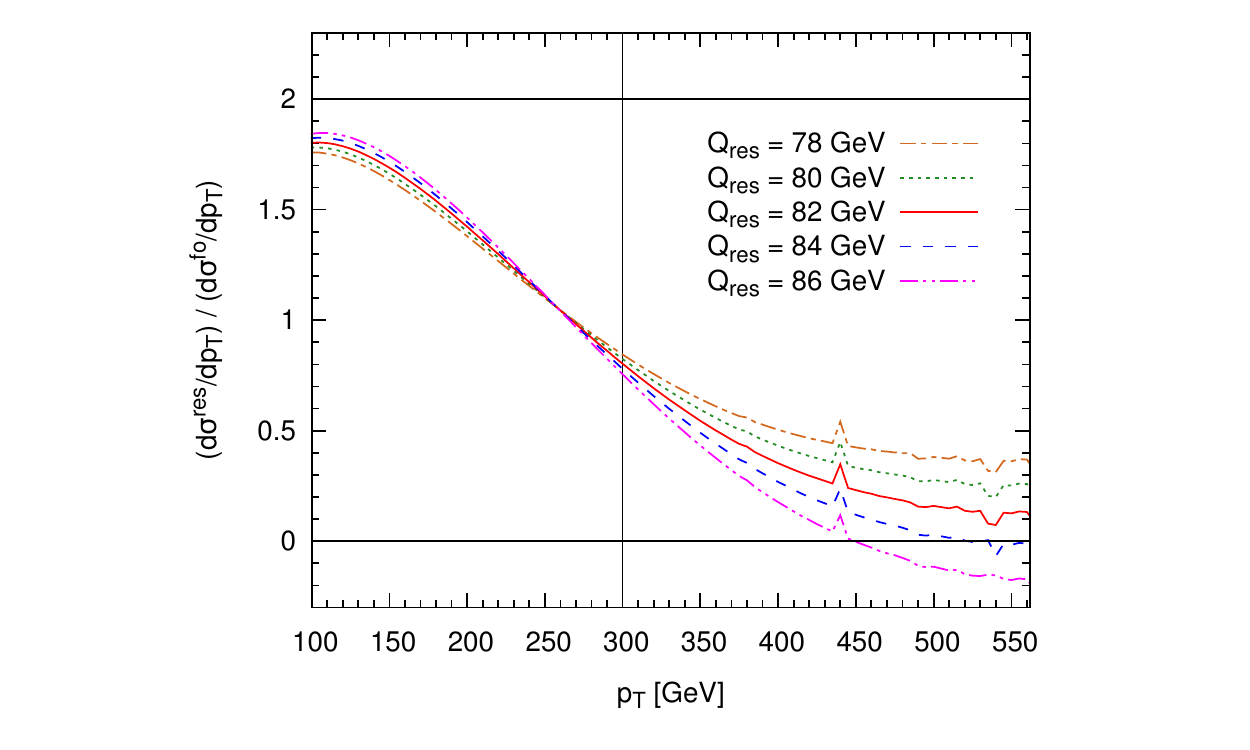}\\[-0.3cm]
    \hspace{-0.7cm} (a) & \hspace{-2cm} (b)\\[0.5cm]
    \multicolumn{2}{c}{%
    \hspace{-1.4cm}
    \includegraphics[angle=0,width=0.64\textwidth]{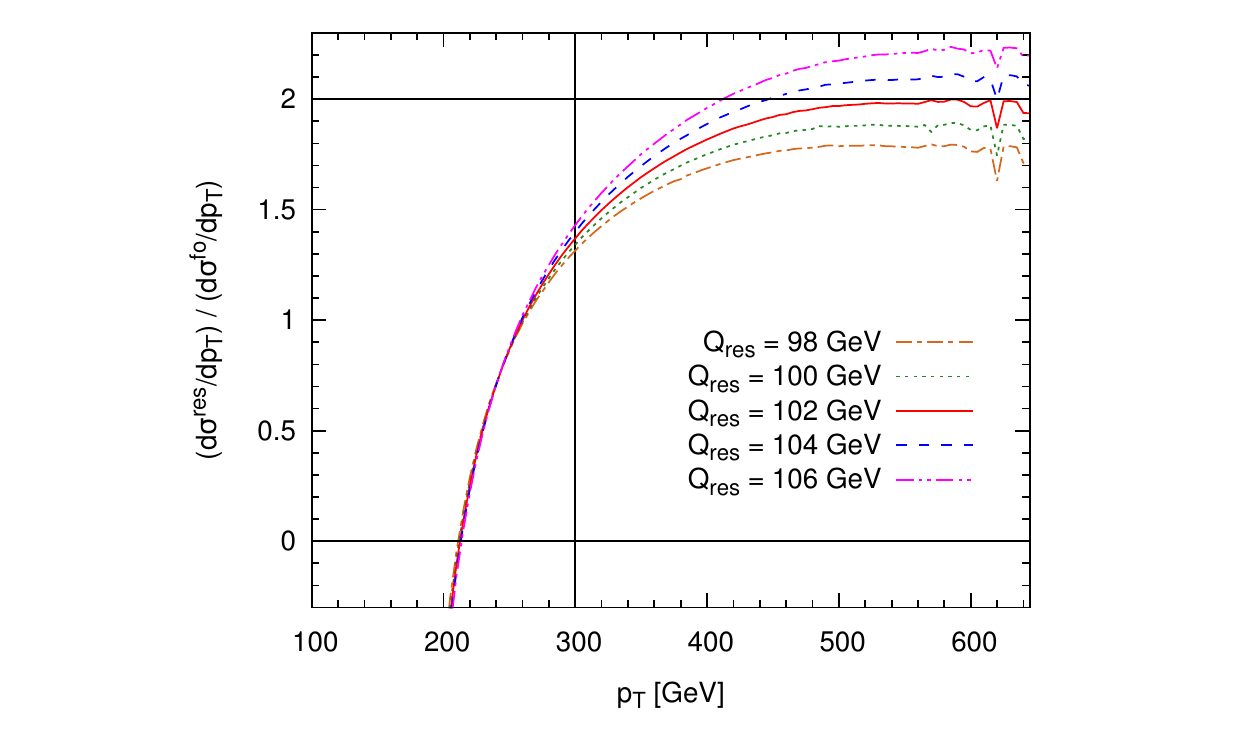}}\\[-0.3cm]
    \multicolumn{2}{c}{\hspace{-0.67cm} (c)}
    \end{tabular}
  \parbox{.9\textwidth}{%
  \caption{\label{fig:qres_h_300}Same as \fig{fig:qres}, but for
    $\mH=300$\,GeV.}}
  \end{center}
\end{figure}

Neglecting squark effects for the moment, we apply this approach
independently to the purely top and bottom induced contributions to the
cross section, as well as to the top-bottom interference term.
\fig{fig:qres} shows these three contributions to the resummed
$p_T$-distribution for different resummation scales in the case of light
Higgs production ($\mh=125.6$\,GeV), normalized to the respective
fixed-order distribution. The curves for a heavy Higgs of $\mH=300$\,GeV
are shown in \fig{fig:qres_h_300}; those for a pseudo-scalar Higgs of
the same mass are very similar to the latter, so we refrain from showing
them here.


\begin{figure}
  \begin{center}
\hspace*{-0.38cm}
    \begin{tabular}{cc}
    \hspace{-1.4cm}
    \includegraphics[angle=0,width=0.64\textwidth]{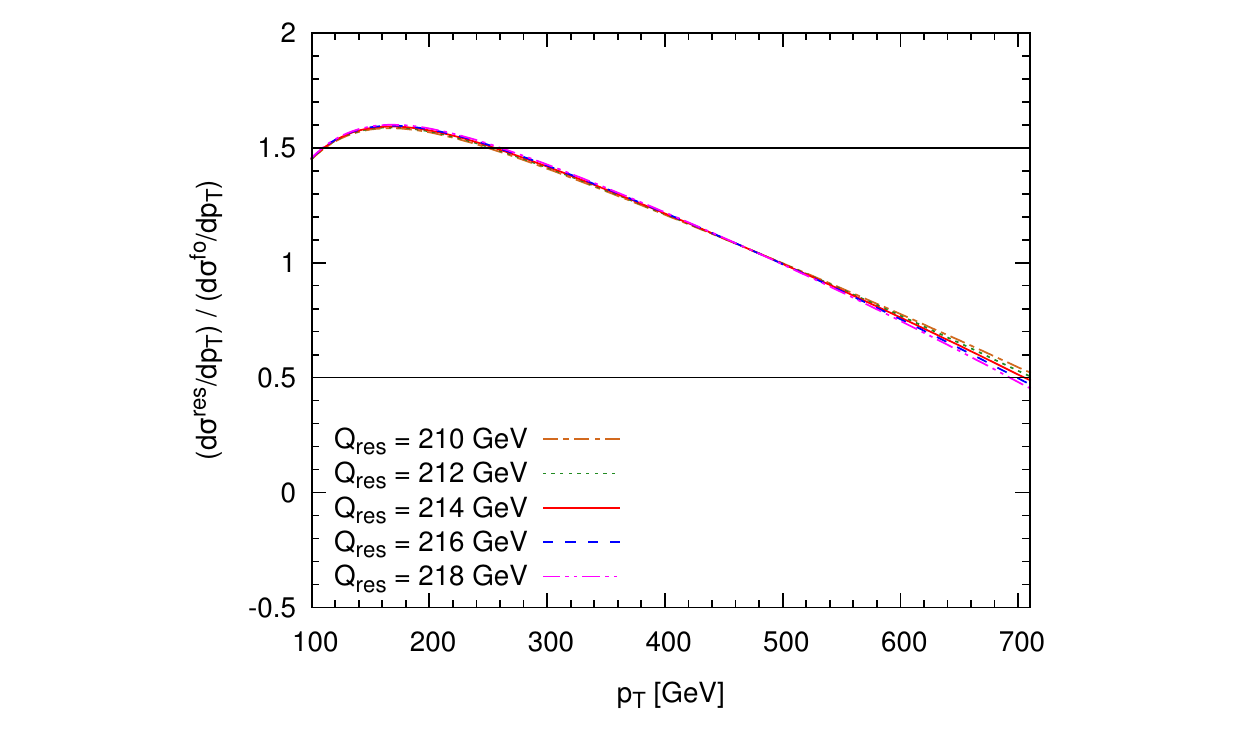}
    &\hspace{-2.73cm}
    \includegraphics[angle=0,width=0.64\textwidth]{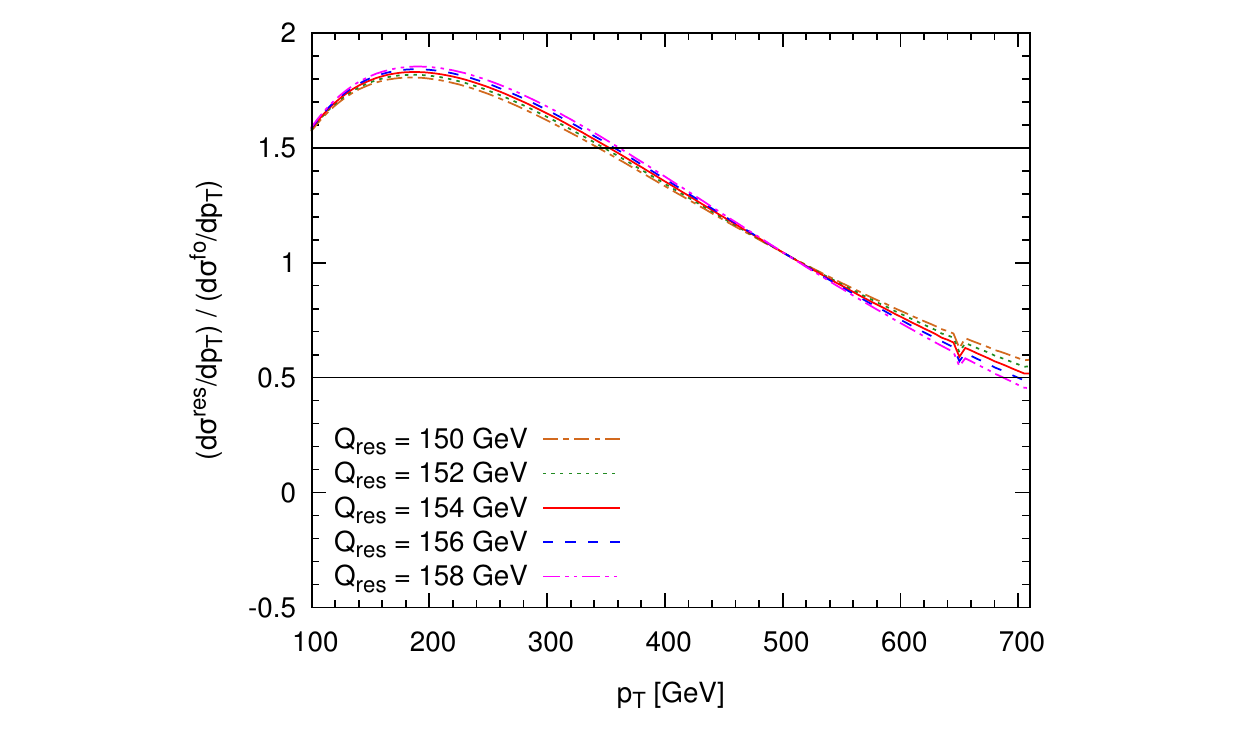}\\[-0.3cm]
    \hspace{-0.7cm} (a) & \hspace{-2cm} (b)\\[0.5cm]
    \multicolumn{2}{c}{%
    \hspace{-1.4cm}
    \includegraphics[angle=0,width=0.64\textwidth]{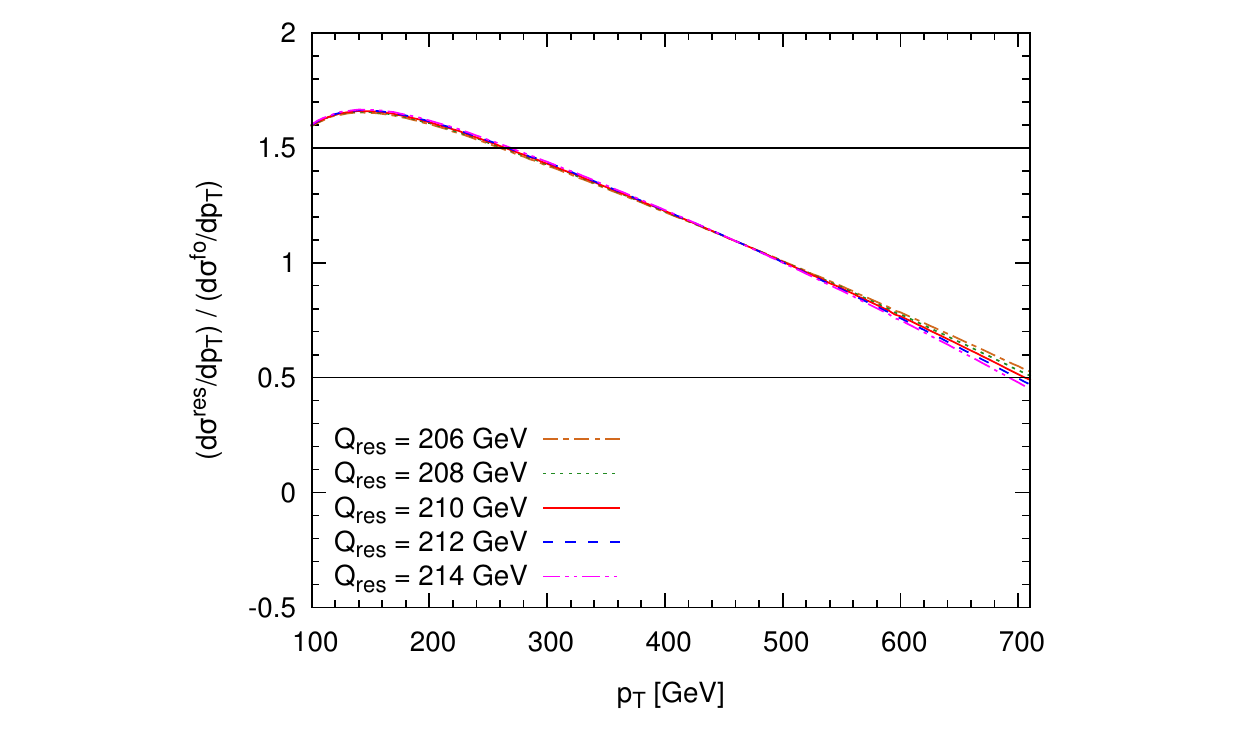}}\\[-0.3cm]
    \multicolumn{2}{c}{\hspace{-0.67cm} (c)}
    \end{tabular}
  \parbox{.9\textwidth}{%
  \caption{\label{fig:qres_h_800}Same as \fig{fig:qres}, but for
    $\mH=800$\,GeV.}}
  \end{center}
\end{figure}

Larger Higgs masses correspond to a harder $\pt$ spectrum since the
larger scale of the process leads to less soft gluon
radiation. Nevertheless, for $\mhiggs=800$\,GeV, the
numerical accuracy of $\dd\sigma^\text{(res)}/\dd\pt^2$ becomes
unreliable already at $\pt\gtrsim 700$\,GeV, so the above procedure for
choosing $\Qres$ cannot be applied.  We are therefore forced to modify
our criterion for $\mhiggs=800$\,GeV; our choice is to require
$|[\dd\sigma^\text{(res)}/\dd\pt^2]/[\dd\sigma/\dd\pt^2]_\text{f.o.}-1|
= 1/2$ at $p_T=700$\,GeV.  The corresponding curves for a heavy Higgs of
$\mH=800$\,GeV are shown in \fig{fig:qres_h_800}; again, those for a
pseudo-scalar Higgs of the same mass are very similar, so we refrain
from showing them here.  


Our {\it central} scale choice is then defined as
$\Qres_0=\Qres^{\text{max}}/2$, while the associated uncertainty is
determined by varying $\Qres$ within the interval $[\Qres_0/2,2\Qres_0]$
(with an additional damping factor for large $\pt$, see \sct{sec:unc}).
The results of this procedure for a hadronic center of mass energy of
$\sqrt{s}=13$\,TeV are listed in \tab{tab:Q0}.  Somewhat reassuringly,
for $\mh=125.6$\,GeV our value for $\Qres_{0,t}$ agrees rather well
with the default choice $\Qres=\mh/2$ of \citere{deFlorian:2011xf}.  On the other
hand, our interval for $Q_{0,b}$ extends to significantly larger values
as the one argued for in \citere{Grazzini:2013mca}. This is even more so
for the interference term for which, in our case, the central
resummation scale is almost the exact average of the $Q_{0,t}$ and
$Q_{0,b}$, while \citere{Grazzini:2013mca} fully attributed this term to
the bottom contribution.


\begin{table}
\begin{center}
  \begin{tabular}{|c|c|c|c|c|}
    \hline
    Higgs type & mass/GeV & $\Qres_{0,t}$/GeV & $\Qres_{0,b}$/GeV &
    $\Qres_{0,\text{int}}$/GeV\\
    \hline\hline
    \multirow{3}{*}{scalar} & 125.6 & 49  & 23  & 34 \\
     & 300 & 62  & 41 & 51 \\
     & 800 & 107  & 77  & 105 \\
    \hline
    \multirow{2}{*}{pseudo-scalar}
     & 300 & 61  & 43  & 49 \\
     & 800 & 117  & 80  & 104 \\
    \hline
  \end{tabular}
  \parbox{.9\textwidth}{%
\caption[]{\label{tab:Q0} Central resummation scales for the top-,
  bottom-, and their interference contribution to the cross section for
  scalar and pseudo-scalar Higgs production at various Higgs masses.}}
\end{center}
\end{table}


Our result for $Q_{0,\text{int}}$ agrees very well with what was found
for the case of jet-veto in \citere{Banfi:2013eda}. By analyzing the
finite remainder of the bottom contribution, which includes the
top-bottom interference in their case, they find $Q\approx 35$\,GeV to
be an appropriate scale choice.\footnote{Note that in the \sm{} the
  bottom contribution is clearly dominated by the top-bottom
  interference term.}

Even though our approach to determine $\Qres$ seems very pragmatic, the
underlying idea is physical, of course. A too large resummation scale
$\Qres$ would overemphasize the Sudakov contribution, typically
overshooting the cross section for $\pt\lesssim\Qres$. Due to the
constraint of \eqn{eq:unitarity}, the only way to compensate for this
effect is to reduce the cross section at larger transverse momenta, such
that it may even become negative. Therefore, by demanding resummation
scales that lead to satisfactory matching at high transverse momenta,
one indirectly restricts resummation to regions where the soft and
collinear factorization is valid. More precisely, we can expect $\Qres$
to be close to the upper boundary of the range allowed by factorization,
certainly not far above that. Note also that there is a certain amount
of freedom how this range is defined.

\begin{figure}
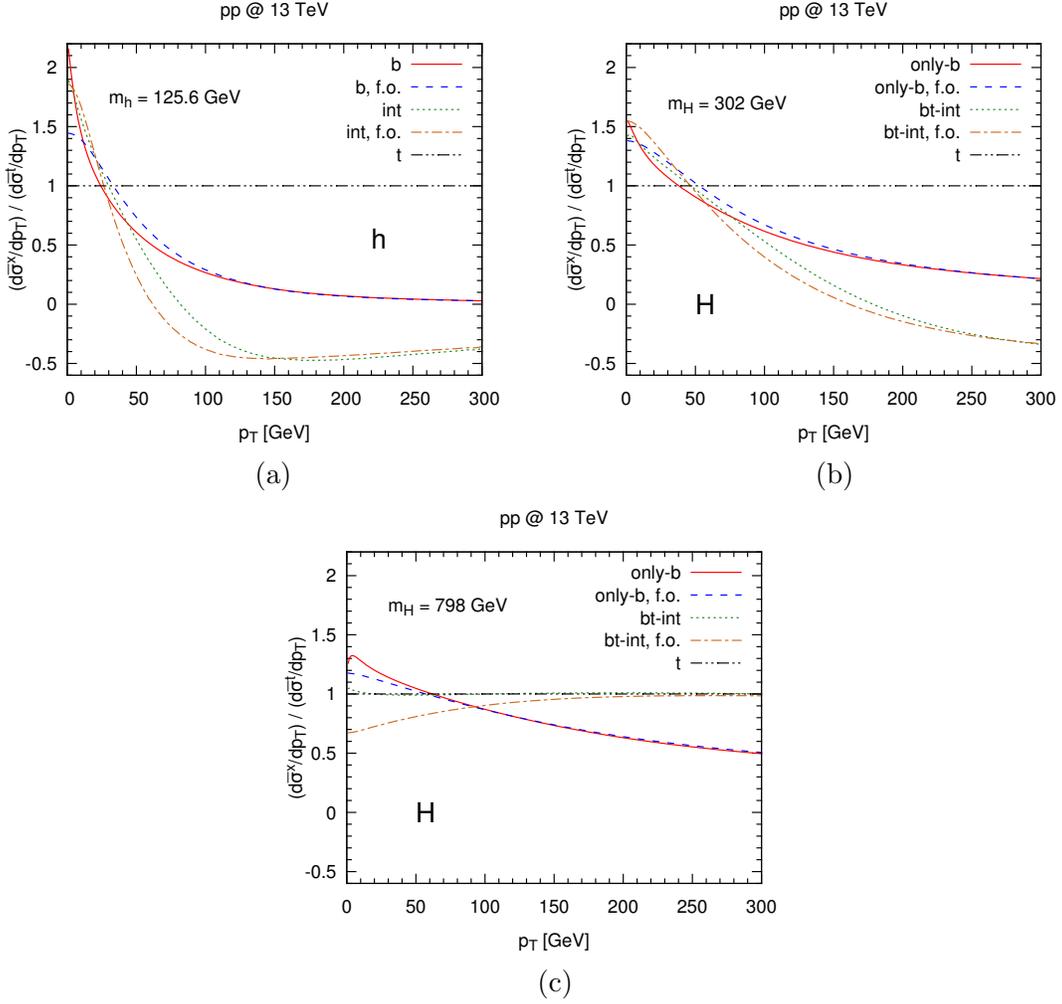

\begin{center}
\hspace{-0.85cm}
\begin{tabular}{cc}
\includegraphics[viewport=70 0 300 210,height=.32\textheight,
  page=17]{plots/pt13.pdf} &
\includegraphics[viewport=70 0 300 210,height=.32\textheight, 
page=18]{plots/pt13h.pdf} \\[-1em]
\hspace{1.5em} (a) & \hspace{2em}(b)\\
\multicolumn{2}{c}{\includegraphics[viewport=70 0 300 210,height=
    .32\textheight, 
page=21]{plots/pt13h.pdf}}\\[-1em]
\multicolumn{2}{c}{\hspace{2em}(c)}
\end{tabular}
  \parbox{.9\textwidth}{%
      \caption[]{\label{fig:tbshape}{\sloppy The $\pt$-shape for the
          bottom-quark (red, solid) and the top-bottom interference
          contribution (green, dotted), normalized to the
          top-contribution (black, dash-double dotted). Resummation
          scales are set as in \tab{tab:Q0}. Also shown are the
          respective ratios for the \nlo{} fixed-order results. (a-c)
          $\mhiggs=125.6/300/800$\,GeV.}}}
\end{center}
\end{figure}

For our later discussion, it will be useful to study the impact of the
top-, bottom- and their interference contribution on the {\it shape} of
the $\pt$ distribution, i.e.
\begin{equation}
\begin{split}
&\frac{1}{\sigma}\frac{\dd\sigma}{\dd\pt} \equiv 
\frac{\dd\bar{\sigma}}{\dd\pt}\,,\qquad
\mbox{with}\quad 
\int\dd \pt\frac{\dd\bar{\sigma}}{\dd\pt} = 1\,.
\label{eq:shape}
\end{split}
\end{equation}
\fig{fig:tbshape} shows this shape of the bottom- and the top-bottom
interference contribution relative to the top contribution, for scalar
and pseudo-scalar Higgs production and three different values of the
Higgs boson mass. In all cases, the bottom-quark distribution is
significantly softer than the top contribution, but the difference
between the two decreases for larger Higgs masses. This behavior is
expected since soft radiation off the quark loop becomes larger for a
smaller quark mass, or, equivalently, larger Higgs
mass\cite{Langenegger:2006wu}.

  The shape of top-bottom interference term experiences a number of
  qualitative and quantitative changes as the Higgs mass increases. For
  $\mhiggs=125.6$\,GeV, it can lead to quite some deviations from a pure
  top- or bottom-dominated shape, see \fig{fig:tbshape}\,(a). Whether
  the spectrum becomes harder or softer depends on the sign of the
  interference (and thus also on the sign of the Yukawa
  couplings).\footnote{ Note that when subtracting a softer spectrum from a
    harder one, the combined spectrum is harder than both of them.}
  Note also that there is a sign change at about $\pt=80$\,GeV.  At
  $\mhiggs=300$\,GeV, \fig{fig:tbshape}\,(b), the qualitative behavior
  remains roughly the same, but appears to be less distinct. At
  $\mhiggs=800$\,GeV, \fig{fig:tbshape}\,(c), on the other hand, the
  shape of the interference term is almost indistinguishable from the
  top contribution, except for very small $\pt$.

  We observe a nice convergence of the resummed and the fixed-order
  distributions towards large $\pt$, as required by our determination of
  the matching scale.  The curves for a pseudo-scalar Higgs are very
  similar to the scalar case which is why we refrain from showing them
  here.


As squark effects are typically small due to the fact that squark masses
are expected to be of the order of a few hundred GeV, we do not determine
separate resummation scales for them. We therefore split the cross section into
three terms:
\begin{itemize}
\item The {\it \bsquare{} contribution} is proportional to the square of
  the bottom-Higgs coupling: $\sigma_\text{\bsquare}\sim y_b^2$. Note
  that $\sigma_\text{\bsquare{}}$ does {\it not} include sbottom effects.
\item The {\it \interference{} contribution} is linearly proportional to
  the bottom Yukawa coupling: $\sigma_\text{\interference}\sim y_b$. It
  therefore contains interference terms of the bottom- with the top- and
  squark-loop induced amplitude.
\item The {\it \tpsquarks{} contribution} is defined as the cross section
  for $y_b=0$ and contains top- and squark-loop induced terms.
\end{itemize}
For the \bsquare{}, the \interference, and the \tpsquarks{}
contribution, we use the resummation scales $Q_{0,b}$,
$Q_{0,\text{int}}$, and $Q_{0,t}$ of \tab{tab:Q0},
respectively, despite the fact that these scales were determined by
disregarding squark effects. We have checked though that the numerical
value of $Q_{0,\text{int}}$ and $Q_{0,t}$ are hardly affected when
squark effects are taken into account.


\section{Input parameters}
\label{sec:input}


We present results for the resummed transverse momentum distribution of
neutral Higgs bosons produced at the \lhc{} via gluon fusion in various
scenarios of the \mssm{}. Our goal here is not a detailed and
comprehensive study of the $\pt$-spectrum in each of these scenarios
though. Rather, we will make use of specific scenarios to highlight
various features and dependences of the $\pt$-spectrum.  Our default
choice for the center-of-mass energy is $13$ TeV. The central
factorization and renormalization scale is set to
$\muF=\muR=\mhiggs{}/2$. The choice for the central resummation scale is
more subtle and is given in \tab{tab:Q0}. All numbers are obtained with
the \nlo{} \pdf{} set of {\abbrev MSTW2008}\cite{Martin:2009iq}, which
implies that the input value for the strong coupling constant is taken
as $\als(m_Z) = 0.12018$. We use the on-shell top and bottom mass with
numerical values $\mt=173.2$\,GeV and $\mb=4.92$\,GeV both for the
internal propagators and the Yukawa couplings.  Terms enhanced by
$\tan\beta$ are implicitely
resummed\cite{Carena:1999py,Guasch:2003cv,Noth:2010jy,%
  Noth:2008tw,Mihaila:2010mp} by reweighing the bottom Yukawa coupling
as described in \citere{Bagnaschi:2014zla}. Similarly, the stop and the
sbottom masses and mixing angles are renormalized as in
\citere{Bagnaschi:2014zla}, in accordance with the definition of the
benchmark scenarios to be described in the next section.


\subsection{\mssm{} parameter points}
\label{sec:scenarios}
We compare results for various \mssm{} benchmark scenarios, as defined
in \citere{Carena:2013qia}\footnote{For the \lightstop{}
  scenario we use the modified version suggested in
  \citere{Bagnaschi:2014zla}.}
and refer to that paper for further details. These benchmark scenarios
require the choice of $\mA$ and $\tan\beta$. Using the exclusion plots
of \citere{Carena:2013qia} and {\tt
  HiggsBounds}~\cite{Bechtle:2008jh,Bechtle:2011sb,Bechtle:2013wla}, we
identified proper (i.e.\ not yet excluded) parameter choices within
the $\mA$-$\tan\beta$ plane, while requiring that $\mh=125.6\pm0.7$\,GeV
(except for the \lightstop{} scenario). The scenarios used for our
analysis in \sct{sec:results} are defined in
\tab{tab:scen}.

\begin{table}
\begin{center}
\begin{tabular}{|c|c|c|c|c|}
\hline
scenario &     $\mA$/GeV & $\tan\beta$ & $\mh$/GeV & $\mH$/GeV \\
\hline
\multirow{2}{*}{\tauphobic} &
    800 & 16 & 125.0 &  798.3\\
  &  800 & 29.5 & 124.9 &  798.4\\
\hline
\multirow{1}{*}{\lightstau} &
    500 & 12 & 125.6 &  500.2\\
\hline
\multirow{3}{*}{ \mhmodp} &
    500 & 17 & 125.6 &  499.9\\
  &  800 & 17 & 125.6 &  800.0\\
  &  800 & 40 & 126.1 &  799.7\\
\hline
\multirow{3}{*}{ \mhmodm} &
    500 & 16.5 & 125.6 &  499.8\\
  &  800 & 16.5 & 125.6 &  800.0\\
  &  800 & 40 & 126.2 &  799.6\\
\hline
\multirow{2}{*}{ \mhmax} &
    300 & 6.5 & 125.7 &  302.1\\
  &  800 & 6.5 & 125.3 &  800.6\\
\hline
\multirow{1}{*}{ \lightstop} &
    800 & 30 & 123.0 &  798.9\\
\hline
\end{tabular}
  \parbox{.9\textwidth}{%
\caption[]{\label{tab:scen} Parameter points considered in this
  paper. The full definition of the scenarios is given in
  \citere{Carena:2013qia}; for the \lightstop{} scenario, however, we
  modify the soft \susy{} breaking wino and bino mass terms as well as
  the $\mu$-parameter as suggested in \citere{Bagnaschi:2014zla}, where
  $m_2=\mu=400$\,GeV and $m_1=340$\,GeV, in order to evade constraints
  on the stop and sbottom masses presented by {\abbrev ATLAS} and
  {\abbrev CMS}. The particular parameter points defined here will be
  refered to in the text as ``scenario($\mA/$GeV,$\tan\beta$)''; for
  example, the first parameter point in the table is \tauphobic(800,16)
  in this notation. The Higgs masses are evaluated with {\tt FeynHiggs} \cite{Heinemeyer:1998yj,Frank:2006yh,Heinemeyer:1998np,Degrassi:2001yf,Brignole:2001jy,Brignole:2002bz,Dedes:2003km,Heinemeyer:2004xw,
  Heinemeyer:2007aq} which we also apply to determine the 
  the corresponding Higgs couplings in the various scenarios.}}
\end{center}
\end{table}


\subsection{Theoretical uncertainties}
\label{sec:unc}

The main sources of theoretical uncertainty on our result for the $\pt$
distribution are due to missing higher order effects, as well as the
uncertainty from the \pdf{}s and $\alpha_s(m_Z)$. The latter two are usually
estimated by following the so-called {\abbrev PDF4LHC} recipe \cite{Botje:2011sn}. 
They will, however, not be part of our analysis within this manuscript.

The former are typically estimated by a variation of unphysical scales
that emerge at finite perturbative or logarithmic order. In our case,
these are the renormalization, the factorization, and the resummation
scale. While for $\muF$ and $\muR$, we follow the standard procedure of
considering the maximum variation of the cross section when
$2\muR/\mhiggs$ and $2\muF/\mhiggs$ are taken from the set
$\{1/2,1,2\}$, while excluding the values for which
$\muR/\muF\in\{1/4,4\}$. The impact of the choice of the resummation
scale, on the other hand, we estimate by varying $\Qres/Q_0$ within the
interval $[1/2,2]$. However, a variation within this
region at large $\pt$ would grossly overestimate the uncertainty at
$\pt\gtrsim \mhiggs$, where the prediction should be well
described by the fixed-order distribution. We therefore modulate the error
band resulting from $\Qres$-variation by a damping factor
\begin{equation}
\begin{split}
d(\pt) = \left[1+\exp(\alpha(\pt-\mhiggs))\right]^{-1}\,,\qquad \alpha =
0.1\,\text{GeV}^{-1}\,,
\label{eq:damp}
\end{split}
\end{equation}
which effectively switches off the $\Qres$-uncertainty for $\pt\gtrsim
\mhiggs$.  Finally, we add the uncertainty estimated from
$\muF$- and $\muR$-variation and the one induced by $\Qres$ in quadrature.



\section{Results}
\label{sec:results}
We are now ready to present our results for the transverse momentum
spectrum of \mssm{} Higgs bosons produced in gluon fusion through
\nlonll.

\begin{figure}
\begin{center}
\includegraphics[trim = 10mm 3mm 0mm 0mm, height=.36\textheight, page=10]{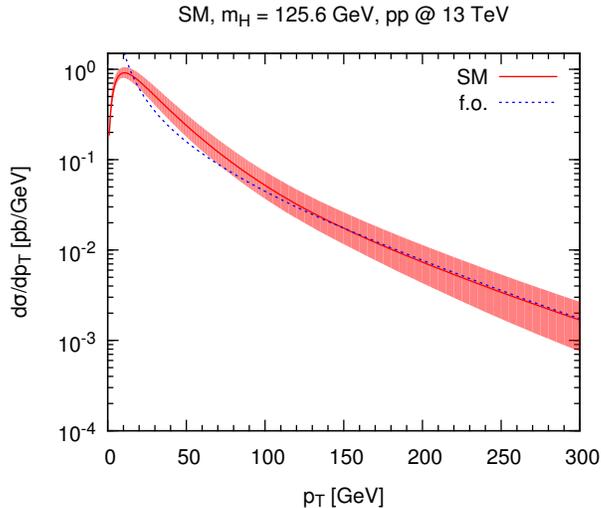}
    \parbox{.9\textwidth}{%
      \caption[]{\label{fig:SM}{\sloppy Transverse momentum distribution
          at \nlo{} (blue, dotted) and \nlo{}\plus\nll{} (red, solid) in
          the \sm{}; lines: central scale choices; band: uncertainty on \nlo{}\plus{}\nll{}
          due to scale variation as described in the text.}  }}
\end{center}
\end{figure}
\begin{figure}
\begin{center}
\includegraphics[trim = 10mm 3mm 0mm 0mm, height=.36\textheight, page=2]{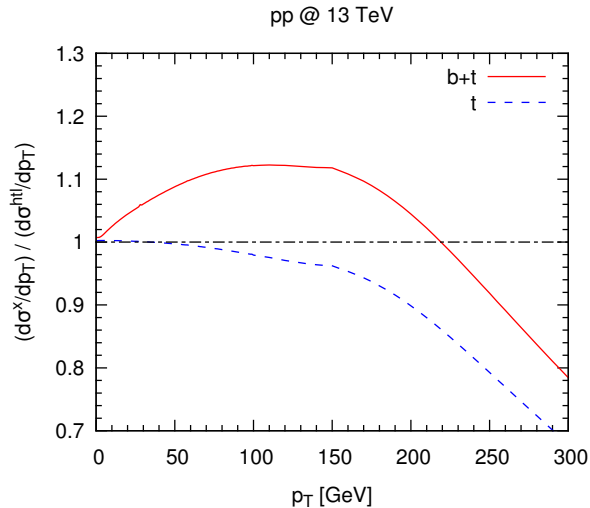}
    \parbox{.9\textwidth}{%
      \caption[]{\label{fig:SMrat}{\sloppy Exact top and bottom mass
          dependence of the transverse momentum distribution at
          \nlonll{}. In the dashed curve, bottom quark effects are set
          to zero. The normalization of these curves is to the
          respective \lo{} total cross section times the \nlonll{}
          result in the heavy top limit.}}}
\end{center}
\end{figure}

b
For reference, \fig{fig:SM} shows the $\pt$-spectrum of a \sm{} Higgs
boson of mass $\mh=125.6$\,GeV.  The resummed cross section is finite in
the limit $\pt{}\rightarrow 0$ and smoothly matches the fixed order
curve at large transverse momenta ($\pt\gtrsim \mh$). The uncertainty
band is obtained through scale variation of $\muF$, $\muR$ and $\Qres$,
following the procedure described in \sct{sec:unc}. In order to compare
to other calculations, it may be useful to consider the ratio of the
$\pt$ distribution which includes the full quark mass dependence to the
result in the heavy-top limit (reweighted by the full \lo{} inclusive
cross section for $gg\to H$). The corresponding curve for the \sm{} is
shown in \fig{fig:SMrat} and can be compared to analoguous plots of
\citeres{Grazzini:2013mca,Bagnaschi:2011tu,Mantler:2012bj,Frixionetalk}.
Disregarding the specific normalization in these papers, the behaviour
of the curve which includes both top- and bottom-quark effects is quite
different in the various approaches, in particular towards small values
of $\pt$. For example, in \citere{Mantler:2012bj}, where a common
resummation scale for the top- and the bottom-quark effects of
$Q_{0,t}=Q_{0,b}=Q_{0,\text{int}}=\mhiggs/2$ was chosen, the curve drops
only by about 6\% between $\pt=100$\,GeV and $\pt=0$. With a separate
resummation scale for the bottom-effects of
$Q_{0,b}=Q_{0,\text{int}}\in[\mb,4\mb]$ as suggested in
\citere{Grazzini:2013mca}, this effect becomes much more pronounced and
amounts to about $(27\pm 9)$\%. In the \powheg{} approach of
\citere{Bagnaschi:2011tu}, on the other hand, the drop in the curve is
roughly 20\%, while the \mcatnlo{}\cite{Frixione:2002ik} result of
\citere{Frixionetalk} with a drop of $5$\% is quite similar to the
analytic resummation.

  Separate resummation scales for the top, the bottom, and the
  interference term as given in \tab{tab:Q0}, on the other hand,
  lead to a drop of 11\%, which is of a size as expected considering the
  magnitude of the new scale choices compared to the ones of the
    previous studies in \citeres{Grazzini:2013mca,Mantler:2012bj}.

   It has been shown that differences in the various approaches
   (analytic resummation, \powheg{}, \mcatnlo{}) become much smaller by
   a simultaneous adjustment of the corresponding intrinsic scales
   ($\Qres$, $h_{\text{fact}}$, shower scale).  Due to the similarities
   in their \nlo{} matching, this leads to an excellent agreement for
   various scale choices \cite{Frixionetalk2} for the analytic
   resummation and \mcatnlo{}. Also, the initially observed large
   differences to the \powheg{} approach are alleviated
   \cite{Vicinitalk} at least when in all approaches the scale for the
   bottom contribution is choosen of the order of the bottom mass.

\begin{figure}
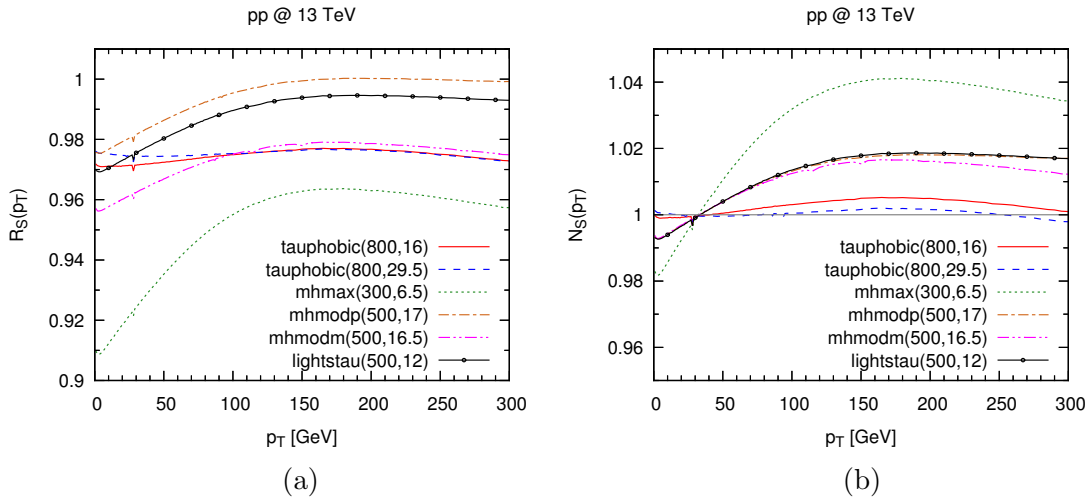

\begin{center}
\hspace{-0.6cm}
\begin{tabular}{cc}
\includegraphics[viewport=70 0 300 210,height=.32\textheight,
  page=11]{plots/pt13.pdf} &
\includegraphics[viewport=70 0 300 210,height=.32\textheight, 
page=13]{plots/pt13.pdf} \\[-1em]
\hspace{1.5em} (a) & \hspace{2em}(b)
\end{tabular}
  \parbox{.9\textwidth}{%
      \caption[]{\label{fig:ratio}{\sloppy Resummed $p_T$ distributions
          in the various scenarios normalized to the \sm{}
          distribution: (a) ratio $R(\scenario{})$ as defined in
          \eqn{eq:R} and (b) $N(\scenario{})$ as defined in
          \eqn{eq:N}.
}}}
\end{center}
\end{figure}

The $\pt{}$ distributions of the light Higgs boson in the various
scenarios of \tab{tab:scen} are virtually indistinguishable from the
\sm{} distribution shown in \fig{fig:SM}. This is because the
observation of a Higgs particle at about $\mh=125$\,GeV typically
constrains the parameter space of the \mssm{} in such a way that the
light Higgs is \sm{}-like.

In order to quantify the deviations between the \mssm{} and the \sm{}
prediction, \fig{fig:ratio}\,(a) shows the ratios
\begin{align}
\label{eq:R}
R_\scenario{}(\pt)=\frac{\dd\sigma_\scenario{}/\dd
    p_T}{\dd\sigma_\text{\sm}/\dd
    p_T}
\end{align}
of the resummed $\pt$ distributions at \nlonll{} for the various \mssm{}
scenarios \scenario{} with respect to the \sm{} one. The difference to
the \sm{} is typically at the 1-3\% level; only scenario \mhmax(300,6.5)
deviates by up to $9$\%.  All curves are below one, because their
respective total cross sections are smaller than the \sm{}
one. Considering the ratio for the shapes, i.e.\ the normalized
distributions, \eqn{eq:shape},
\begin{align}
\label{eq:N}
N_\scenario{}(\pt)=\frac{\dd\bar{\sigma}_\scenario{}/\dd
  p_T}{\dd\bar{\sigma}_\text{\sm}/\dd p_T}\,,
\end{align}
we find variations at the 2\%-level, see \fig{fig:ratio}\,(b), with the
only exception again \mhmax(300,6.5) which, however, still stays within
4\% of the \sm{} prediction. Apparently, the harder spectrum for the
\mssm{} scenarios compared to the \sm{} is due to a slightly larger
negative \interference{} term, recall \fig{fig:tbshape}. The numerical
effects observed here are roughly of the same size as those observed in
\citere{Bagnaschi:2011tu} (see the right plot of Fig.\,8 in that paper).
The fact that for all scenarios $\scenario$, the ratio $N_\scenario = 1$
occurs at roughly the same value of $\pt\approx 30$\,GeV is a
consequence of the similar ``barycenter''
$\langle\hat{p}_T\rangle=\sigma_\text{tot}^{-1}\int\dd\pt\,\pt\,
(\dd\sigma/\dd\pt)$ of the distributions.

\begin{figure}
\begin{center}
    \begin{tabular}{cc}
     \hspace{-0.45cm}\mbox{\includegraphics[trim = 15mm 3mm 0mm 0mm,
         height=.295\textheight, page=1]{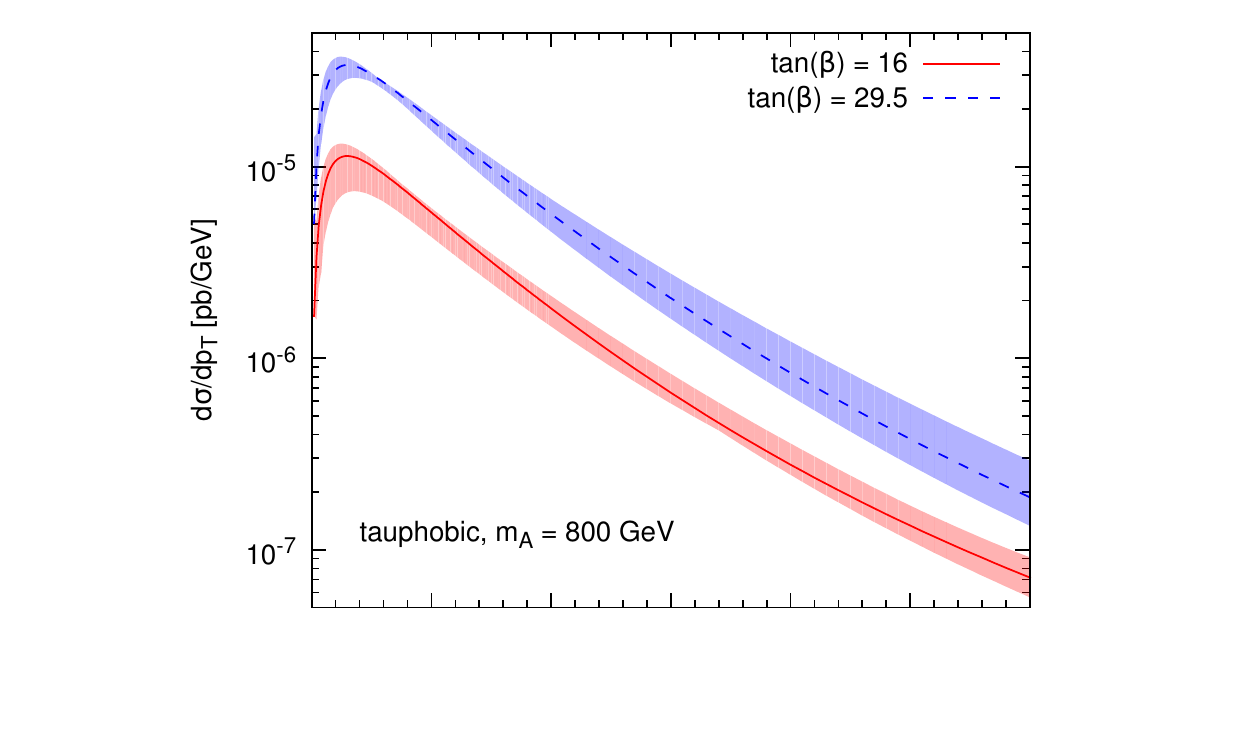}}
     & \hspace{-1.8cm}\mbox{\includegraphics[trim = 15mm 3mm 0mm 0mm,
         height=.295\textheight, page=1]{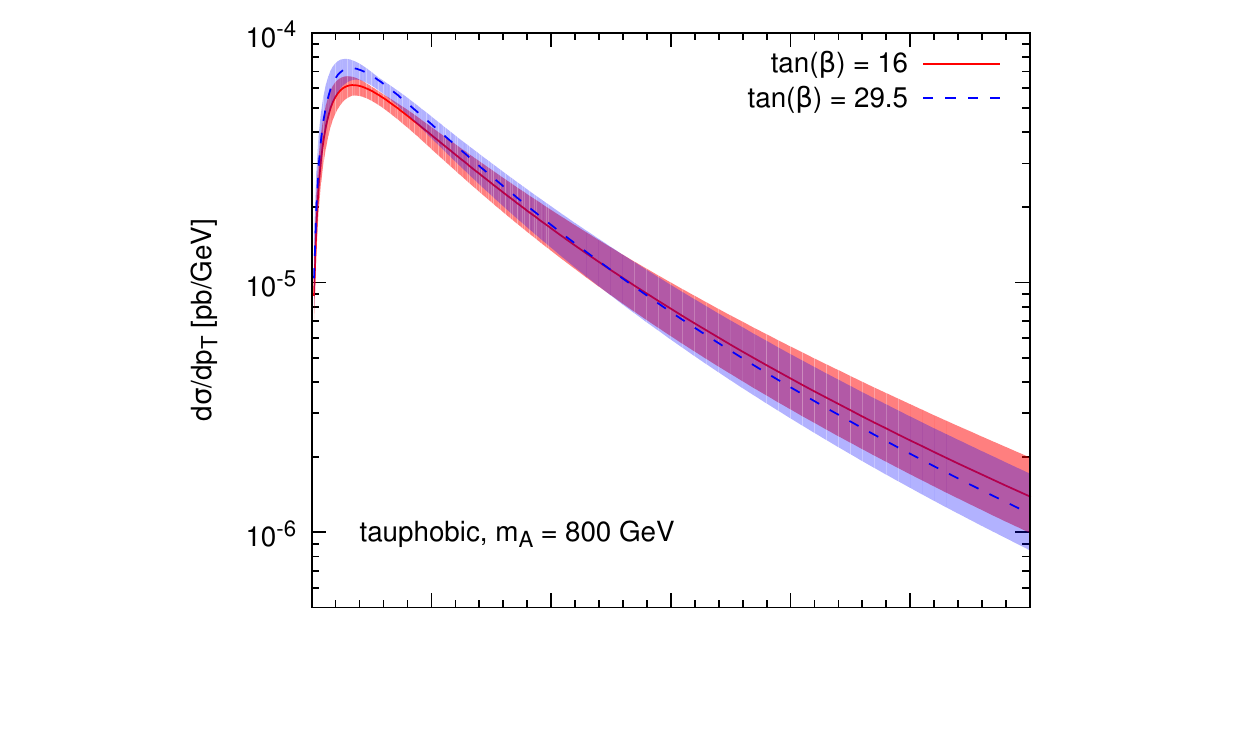}} \\[-1.365cm]
     \hspace{-0.45cm}\mbox{\includegraphics[trim = 15mm 3mm 0mm 0mm,
         height=.295\textheight, page=2]{plots/pt13h.pdf}}
     & \hspace{-1.8cm}\mbox{\includegraphics[trim = 15mm 3mm 0mm 0mm,
         height=.295\textheight, page=2]{plots/pt13p.pdf}} \\[-1.365cm]
     \hspace{-0.45cm}\mbox{\includegraphics[trim = 15mm 3mm 0mm 0mm,
         height=.295\textheight, page=3]{plots/pt13h.pdf}}
     & \hspace{-1.8cm}\mbox{\includegraphics[trim = 15mm 3mm 0mm 0mm,
         height=.295\textheight, page=3]{plots/pt13p.pdf}} \\
     \hspace{-1cm} (a) & \hspace{-2.4cm}(b)
    \end{tabular}
    \parbox{.9\textwidth}{%
      \caption[]{\label{fig:pThp}{\sloppy Transverse momentum
          distribution at \nlo{}\plus\nll{} for (a) the heavy and (b)
          the pseudo-scalar \mssm{} Higgs boson for scenarios
          \tauphobic(800,16) (red, solid), \tauphobic(800,29.5) (blue,
          dashed), \mhmodp(800,17) (green, dotted), \mhmodp(800,40)
          (magenta, dash-double dotted), \mhmodm(800,16.5) (black,
          solid with dots) and \mhmodm(800,40) (brown, dash-dotted);
          lines: central scale choices; bands: uncertainty due to scale
          variation.  } }}
\end{center}
\end{figure}

While the predictions for the light Higgs are very \sm{}-like, this is
not the case for the heavy and pseudo-scalar Higgs, see
\fig{fig:pThp}.\footnote{In fact, note that, when the light Higgs is
  close to the decoupling limit, i.e. its couplings become identical to
  the \sm{} ones, the opposite is the case for the heavy Higgs.} We show
curves for various scenarios with $\mA=800$ GeV, where $\mH\approx \mA$.
Clearly, the absolute size of the cross section for both $H$ and $A$
depends strongly on the respective scenario and the value of $\tan\beta$
(see also \citere{Bagnaschi:2014zla}).  Indeed for the heavy Higgs, in
all scenarios the cross section increases with the value of $\tan\beta$,
which is caused by the fact that the bottom contribution strongly increases
and eventually becomes the dominant contribution to the cross section.  One
remarkable observation is that for the pseudo-scalar Higgs the curves in
the \tauphobic{} scenarios for the two different values of $\tan\beta$
are quite close, in contrast to all other scenarios. In fact, at large
values of $\pt{}$, the cross section for $\tan\beta=16$ is even bigger
than the one for $\tan\beta=29.5$. The reason for this behavior will be
discussed further below.


\begin{figure}
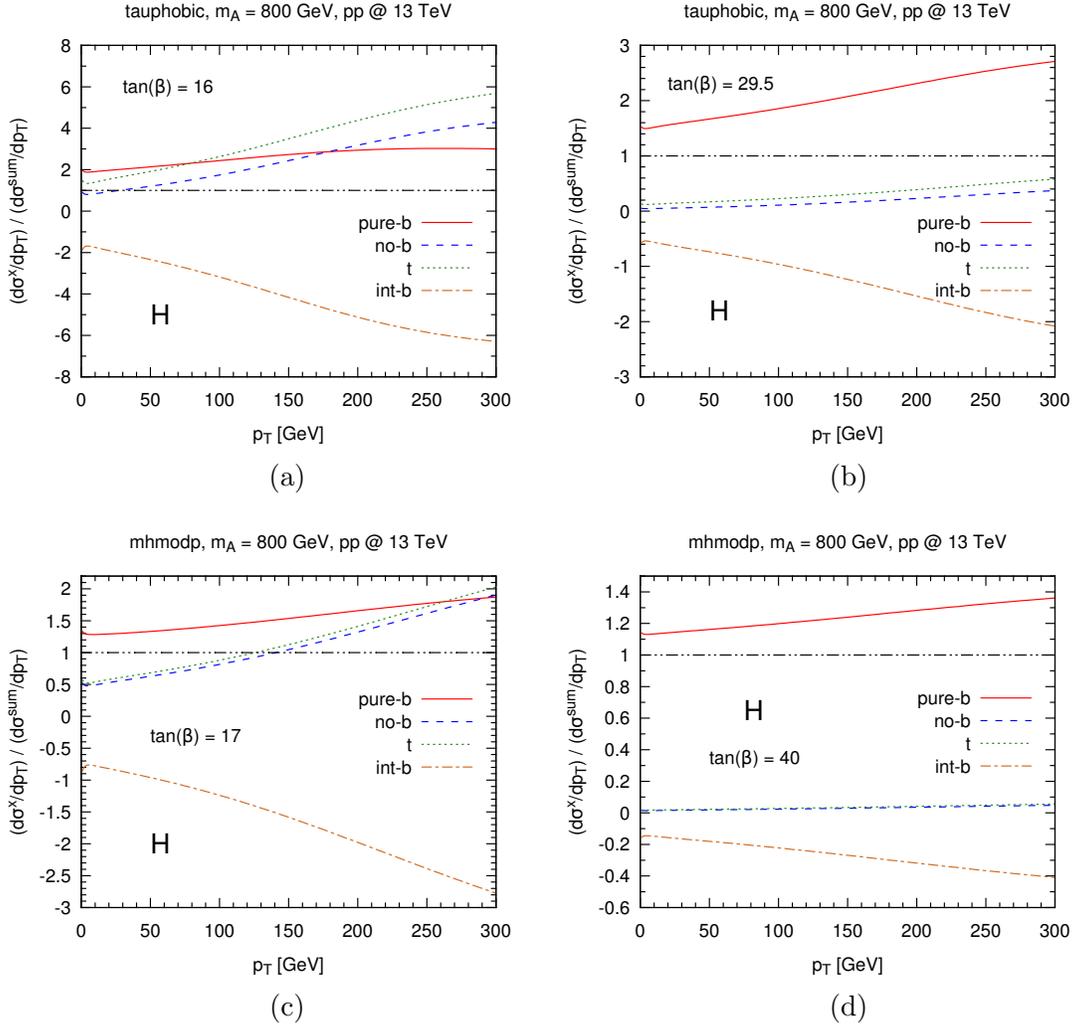

\begin{center}
\hspace{-0.85cm}
\begin{tabular}{cc}
\includegraphics[viewport=70 0 300 210,height=.32\textheight,
  page=4]{plots/pt13h.pdf} &
\includegraphics[viewport=70 0 300 210,height=.32\textheight, 
page=5]{plots/pt13h.pdf} \\[-1em]
\hspace{1.5em} (a) & \hspace{2em}(b)\\[0.3cm]
\includegraphics[viewport=70 0 300 210,height=.32\textheight,
  page=10]{plots/pt13h.pdf} &
\includegraphics[viewport=70 0 300 210,height=.32\textheight, 
page=11]{plots/pt13h.pdf} \\[-1em]
\hspace{1.5em} (c) & \hspace{2em}(d)
\end{tabular}
  \parbox{.9\textwidth}{%
      \caption[]{\label{fig:relS}{\sloppy Relative contributions to the
          resummed $p_T$ distribution for the heavy Higgs normalized to
          the full cross section in (a) \tauphobic(800,16), (b)
          \tauphobic(800,29.5), (c) \mhmodp(800,17) and (d)
          \mhmodp(800,40). The pure bottom term (red, solid), the
          \tpsquarks{} term (blue, dashed) and their interference
          (brown, dash-dotted) add up to one, which is marked for reference
          (black, dash-double dotted). For comparison, the pure top
          contribution is shown as well (green, dotted).  } }}
\end{center}
\end{figure}


In \sct{sec:setres}, we split the cross section into the three
contributions \bsquare{}, \tpsquarks{}, and \interference{} for which
separate resummation scales were determined.  The relative contribution
of these three terms to the heavy Higgs $\pt{}$ distribution for (a)~the
\tauphobic(800,16) and (b)~the \tauphobic(800,29.5) scenario is shown in
\fig{fig:relS}. Note that by definition the \bsquare{} (red, solid),
\tpsquarks{} (blue, dashed) and \interference{} curve (brown, dash-dotted)
add up to one (black, dash-double dotted). For comparison, we also
include a curve for the ``\tsquare{} contribution'' (green,
dotted) which is defined to be proportional to the square of the
top-Higgs coupling $y_t$.

For \tauphobic(800,16), we find a rather large cancellation between the
positive \tpsquarks- and \bsquare-, and the negative
\interference{} term, see \fig{fig:relS}\,(a). It shows the importance
of the proper treatment of the \interference{} term in the resummation
procedure and justifies a separate resummation scale as introduced in
\sct{sec:setres}. By comparing to the \tsquare{} contribution, we also
observe that the squark effects are of the order of the overall
contribution and therefore very relevant.

Going to $\tan\beta=29.5$, the cancellations among the individual
contributions are less severe, see \fig{fig:pThp}\,(b). As expected, the
cross section is largely dominated by bottom-quark effects, i.e., the
\bsquare{} and the \interference{} term. These results
substantiate that the contribution of the bottom loop
causes the increase of the cross section that we observed in
\fig{fig:pThp}\,(a) at high $\tan\beta$.


\begin{figure}
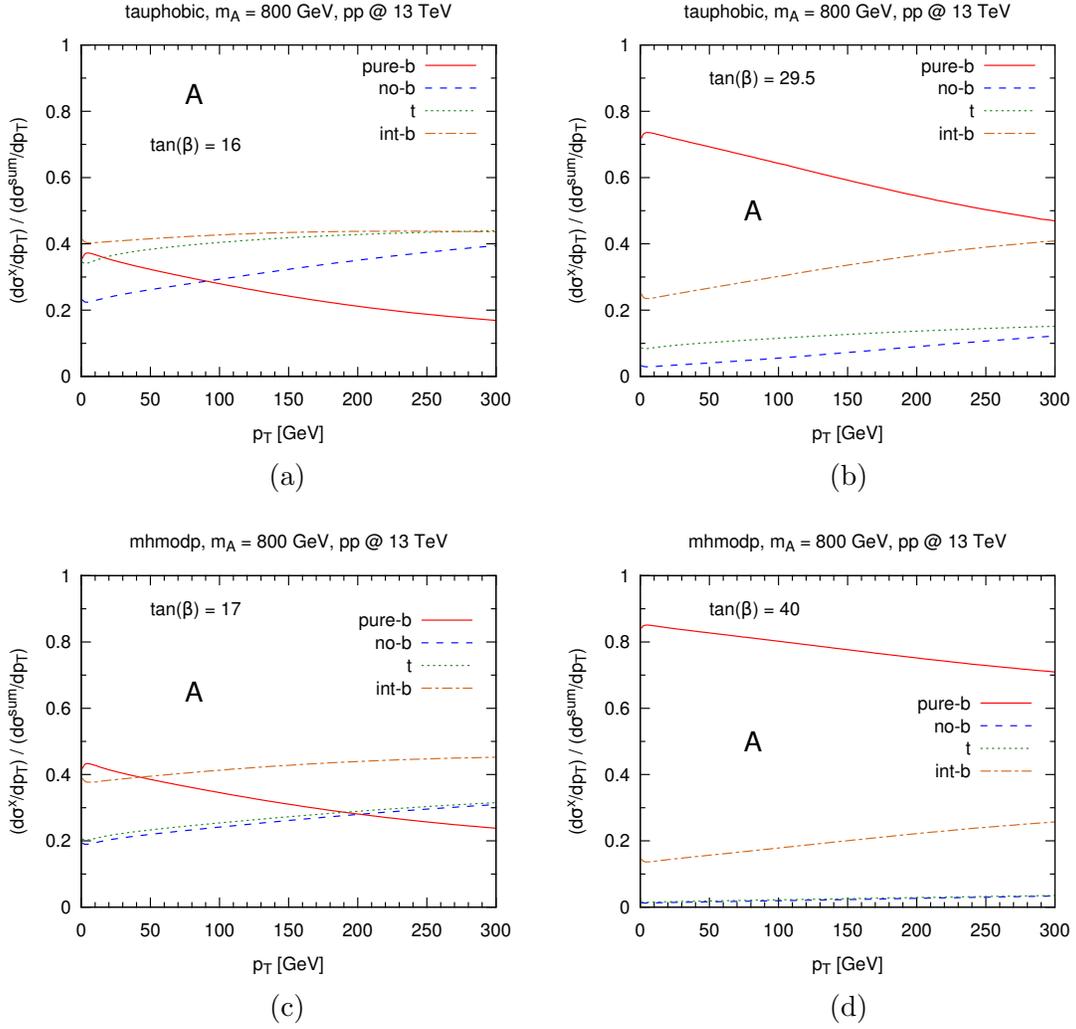

\begin{center}
\hspace{-0.85cm}
\begin{tabular}{cc}
\includegraphics[viewport=70 0 300 210,height=.32\textheight,
  page=4]{plots/pt13p.pdf} &
\includegraphics[viewport=70 0 300 210,height=.32\textheight, 
page=5]{plots/pt13p.pdf} \\[-1em]
\hspace{1.5em} (a) & \hspace{2em}(b)\\[0.3cm]
\includegraphics[viewport=70 0 300 210,height=.32\textheight,
  page=10]{plots/pt13p.pdf} &
\includegraphics[viewport=70 0 300 210,height=.32\textheight, 
page=11]{plots/pt13p.pdf} \\[-1em]
\hspace{1.5em} (c) & \hspace{2em}(d)
\end{tabular}
  \parbox{.9\textwidth}{%
      \caption[]{\label{fig:relSp}{\sloppy Same as \fig{fig:relS}, but
          for the pseudo-scalar Higgs.  } }}
\end{center}
\end{figure}


Let us now compare these observations in the \tauphobic{} to the
\mhmodp{} scenario shown in \fig{fig:relS}\,(c) and (d) (the curves for
the \mhmodm{} scenario are almost indistinguishable from \mhmodp{}). The
qualitative features of the plots in the two scenarios are quite
similar. However, in the \mhmodp{} scenario, the cancellation between
the \interference- and the other terms is less pronounced. Also, the
\bsquare-contribution is typically more important than the \tpsquarks{}
one in this scenario, except for the large-$\pt$ region in the case
$\tan\beta=17$. Squark effects are much smaller there, and at
$\tan\beta=40$, the \tpsquarks-contribution is basically
negligible over the full $\pt$ range.

\fig{fig:relSp} shows the same study for the pseudo-scalar Higgs (again,
\mhmodm{} is almost identical to \mhmodp). The structure of the
various contributions to the cross section is quite different from the
one for the heavy Higgs. Since the \interference{} term is positive here,
all curves remain between $0$ and $1$. In both scenarios, for the
smaller value of $\tan\beta$, each term contributes at least about 20\%
to the cross section, and none of them exceeds 45\%.  The largest
contribution is due to the \interference{} term for most transverse
momenta. At high $\tan\beta$, the \bsquare{} contribution becomes
clearly dominant again. While the \interference{} term remains sizable,
both \tsquare{} and \tpsquarks{} terms are negligible, especially in
the \mhmodp{} scenario.

With the results of \fig{fig:relSp}\,(a) and (b), it is interesting to
take another look at the behavior of the distribution for the
pseudo-scalar Higgs shown in \fig{fig:pThp}\,(b) in the \tauphobic{}
scenarios. The splitting into the individual contributions suggests that
there is no deep reason for the similarity of the curves for
$\tan\beta=16$ and $\tan\beta=29.5$. The hierarchy of the various
contributions to the cross section in the \tauphobic{} scenarios is not
very different from the \mhmodp{} scenarios of \fig{fig:relSp}. It
rather seems to be an accidental interplay of the top- and bottom-quark
effects so that the absolute size of the increase of the
\bsquare{} contribution from $\tan\beta=16$ to $\tan\beta=29.5$ is
compensated by a decrease of similar size of the \tpsquarks{} and
\interference{} term.

\begin{figure}
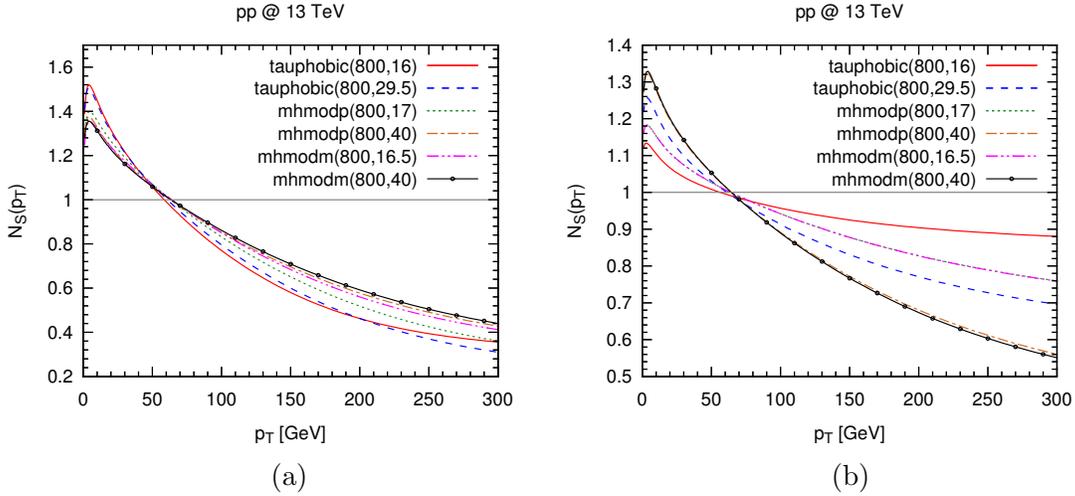

\begin{center}
\hspace{-0.85cm}
\begin{tabular}{cc}
\includegraphics[viewport=70 0 300 210,height=.32\textheight,
  page=15]{plots/pt13h.pdf} &
\includegraphics[viewport=70 0 300 210,height=.32\textheight, 
page=15]{plots/pt13p.pdf} \\[-1em]
\hspace{1.5em} (a) & \hspace{2em}(b)
\end{tabular}
  \parbox{.9\textwidth}{%
      \caption[]{\label{fig:ratioHA}{\sloppy The ratio $N(\scenario)$,
          defined in \eqn{eq:N}, of the resummed $p_T$ shapes in
          \fig{fig:pThp} for (a) the heavy Higgs and (b) the
          pseudo-scalar Higgs.  }}}
\end{center}
\end{figure}

To finalize the analysis of the $p_T$ distributions shown in
\fig{fig:pThp}, we study their shapes in the various scenarios for the
heavy Higgs in \fig{fig:ratioHA}\,(a) and for the pseudo-scalar Higgs in
\fig{fig:ratioHA}\,(b) by considering again the ratio of shapes defined
in \eqn{eq:N}.
Note that the normalization is for a ``\sm{} Higgs'' of mass $800$\,GeV.
For the heavy Higgs in \fig{fig:ratioHA}\,(a), we observe generally
small deviations between the curves.  While the biggest difference
occurs at large transverse momenta, the similarity in shape at small
$p_T$ is remarkable. Their deviation for the \sm{} curve is quite large
though, reaching up to 60\%, and clearly showing the dominance of the
\bsquare{} term by the significantly softer spectrum, see
\fig{fig:tbshape}.  For most scenarios, however, the softness decreases
with increasing $\tan\beta$, with the exception of the \tauphobic{}
scenarios. This is again the impact of the negative \interference{}
term. The softening of the spectrum due to an enhanced $b$-contribution
is in agreement with the observations of
\citeres{Bagnaschi:2011tu,Langenegger:2006wu}.

Considering the pseudo-scalar Higgs in \fig{fig:ratioHA}\,(b), the
spread of the curves is significantly larger both at small and high
transverse momenta, leading to a more enhanced difference in shape of
the resummed $p_T$ distributions in the various scenarios. Since
  the interference contributions are strictly positive in this case, the
  corrections for all scenarios, including \tauphobic{}, decrease with
  increasing $\tan\beta$. We also note that the
$\mh^{\text{mod}+}$ and $\mh^{\text{mod}-}$ curves are practically
indistinguishable in this case.

\begin{figure}[h]
\begin{center}
\includegraphics[trim = 10mm 3mm 0mm 0mm, height=.36\textheight, page=7]{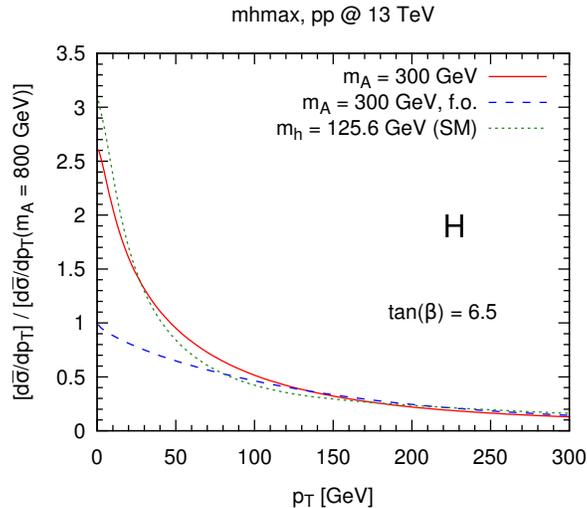}
    \parbox{.9\textwidth}{%
      \caption[]{\label{fig:mH}{\sloppy Ratio of resummed
          $\pt$ shapes in the $\mh^{\text{max}}$ scenario at
          $\mH=300$ GeV and $\mH=800$ GeV (red, solid) for the heavy
          Higgs, the corresponding fixed order curve (blue, dashed)
          and, for comparison, the ratio of the resummed distribution of
          the light Higgs at $\mh=125.6$ GeV and the heavy at $\mH=800$
          GeV (green, dotted).}  }}
\end{center}
\end{figure}

As a final study, we compare the transverse momentum distributions at
different Higgs masses. \fig{fig:mH} shows the ratio of the shapes for
the heavy scalar in the \mhmax{(300,6.5)} and in the \mhmax{(800,6.5)}
scenario. For comparison, the same ratio is shown at fixed order (blue,
dashed). In addition, the ratio of the $\pt$-shape of a \sm{} Higgs at
$125.6$\,GeV and the heavy Higgs in the \mhmax{(800,6.5)} scenario
(green, dotted) is given. The spectra at low Higgs masses are
significantly softer due to increased soft radiation.  This observation
is consistent with the behavior of \figs{fig:SM} and
\ref{fig:pThp}\,(a), which differ by an order of magnitude in the
difference between the minimum at $p_T=300$ GeV and the maximum of the
curve. Furthermore, \fig{fig:mH} confirms that the shape at high
transverse momenta is driven by the fixed order cross section, as
expected. The harder spectrum at high Higgs masses is caused by the fact
that the colliding gluons carry more energy in a production of a heavy
particle, which makes it more likely to emit harder gluons and
therefore, to produce a harder Higgs boson.


\section{Conclusions}
\label{sec:conclude}

We have presented typical \mssm{} effects on the \pt{} spectrum of a
neutral Higgs boson produced at the \lhc{}. Special emphasis has been
put on the impact of bottom quarks. While the current experimental data
imply a small, \sm{}-like bottom-Yukawa coupling for the light Higgs
boson, the production mechanism for the heavy and the pseudo-scalar
Higgs can be dominated by bottom-quark loops.

Through a pragmatic argumentation based on simple theoretical and
physical expectations, we derived separate resummation scales for the
\bsquare{} and \interference{} contributions which turn out to be
significantly larger than the bottom-quark mass, and smaller than the
Higgs boson mass.

We find the well-known behavior that the bottom loop typically softens
the $\pt$-spectrum. For the pseudo-scalar Higgs boson, the
\interference{} term is typically positive and leads to a further
softening of the spectrum.  For the light and heavy {\abbrev CP}-even
Higgs bosons, on the other hand, the \interference{} term is typically
negative, and makes the spectrum a bit harder. The latter effect is
specifically visible for the light Higgs where the spectrum in all
\mssm{} benchmarks is harder than in the \sm{}, because the relative
contribution of the negative \interference{} term is larger. In all
scenarios, we observe an enhanced importance of the bottom-quark
contributions for the heavy and pseudo-scalar Higgs, which become by far
dominant at large values of $\tan\beta$. Indeed, the corresponding
spectra are clearly softer than in the \sm{}.  Finally, we confirmed
that larger Higgs masses lead to reduced soft gluon emission and
therefore a harder spectrum.

The resummed $\pt$-distributions through \nlonll{} have been implemented
in the program {\tt MoRe-SusHi}, which advances the program {\tt SusHi} to
small-$\pt$ distributions. The code is publically available and can be
found on the {\tt SusHi}
homepage.\footnote{\url{http://sushi.hepforge.org/moresushi}}

As an outlook, one may consider combining the consistent \nlonll{}
results presented in this paper for the $\pt$-distribution of \mssm{}
Higgs bosons with the \nnlo{}\plus{}\nnll{} distribution in the
infinite-top mass approximation, and possibly even with the \pt{}
spectrum of the Higgs boson produced through bottom-quark
annihilation. This is certainly beyond the scope of this paper and is
left for a future publication.


\paragraph{Acknowledgements.}
\label{sec:ack}
 We are indebted to Stefan Liebler for comments on the manuscript, for
 suggesting the name ``{\tt MoRe-SusHi}'', and for letting us use the
 diagrams in \fig{fig:diag}. We would like to thank Alessandro Vicini
 for helpful communication and comments on the manuscript. MW would like
 to thank Pier Francesco Monni for several fruitful discussions.  This
 research was supported by the Munich Institute for Astro- and Particle
 Physics (MIAPP) of the DFG cluster of excellence ``Origin and Structure
 of the Universe'' and the BMBF, contract 05H12PXE.  HM was supported by
 a Marie Curie Early Initial Training Network Fellowship of the European
 Community's Seventh Framework Programme under contract number
 (PITN-GA-2012-315877-MCnetITN). MW was supported by the European
 Commission through the FP7 Marie Curie Initial Training Network
 ``LHCPhenoNet'' (PITN-GA-2010-264564).


\appendix
\gdef\thesection{Appendix \Alph{section}}

\section{\bld{g^{(n)}}-functions at \nll{}}
\label{app:gs}
The Sudakov form factor can be expressed in terms of functions $g^{(n)}$, when the argument of the exponential is expanded with respect to $\als$, while treating $\als L$ as being of order unity, see \eqn{eq:sudakov}. To achieve \nll{} accuracy the first two functions have to be taken into account. Their functional expressions for $\Qres=\muR=M$ are the following:
\begin{align}
\begin{split}
g^{(1)}(\als L)&=\frac{A_c^{(1)}}{\beta_0}\frac{\lambda+\ln(1-\lambda)}{\lambda},\\
g^{(2)}(\als L)&=\frac{B_c^{(1)}}{\beta_0} \ln(1-\lambda)-\frac{A_c^{(2)}}{\beta_0^2}\left(\ln(1-\lambda)+\frac{\lambda}{1-\lambda}\right)\\
&+\frac{A_c^{(1)}\beta_1}{\beta_0^3}\left(\frac12\ln^2(1-\lambda)+\frac{\lambda+\ln(1-\lambda)}{1-\lambda}\right),
\end{split}
\end{align}
where 
\begin{align}
\lambda=\beta_0\frac{\als(M)}\pi L
\end{align}
and $\beta_0 = (11\,C_A - 2\,N_f)/12$ and
$\beta_1=(17\,C_A^2-5\,C_A\,N_f-3\,C_F\,N_f)/24$ are the first two
coefficients of the $\beta$ function.

 The process and resummation scheme independent
coefficients needed at \nll{} have been known for some time for gluon-induced processes
\cite{Catani:1988vd}; they read
\begin{equation}
\begin{split}
\label{eq:rescoeff1}
A^{(1)}_g &= C_A\,,\qquad
 A^{(2)}_g = \frac{1}{2} \,C_A
\left[ 
\left(\frac{67}{18} -\frac{\pi^2}{6} \right)C_A -\frac{5}{9} N_f      
\right]\,,\\
B^{(1)}_g &= -\beta_0=-\left(\frac{11}6 C_A+\frac13 N_f\right)\,,\\
\end{split}
\end{equation}
where $C_F=4/3$, $C_A=3$, and $N_f=5$ is the number of active quark
flavors.


\def\app#1#2#3{{\it Act.~Phys.~Pol.~}\jref{\bf B #1}{#2}{#3}}
\def\apa#1#2#3{{\it Act.~Phys.~Austr.~}\jref{\bf#1}{#2}{#3}}
\def\annphys#1#2#3{{\it Ann.~Phys.~}\jref{\bf #1}{#2}{#3}}
\def\cmp#1#2#3{{\it Comm.~Math.~Phys.~}\jref{\bf #1}{#2}{#3}}
\def\cpc#1#2#3{{\it Comp.~Phys.~Commun.~}\jref{\bf #1}{#2}{#3}}
\def\epjc#1#2#3{{\it Eur.\ Phys.\ J.\ }\jref{\bf C #1}{#2}{#3}}
\def\fortp#1#2#3{{\it Fortschr.~Phys.~}\jref{\bf#1}{#2}{#3}}
\def\ijmpc#1#2#3{{\it Int.~J.~Mod.~Phys.~}\jref{\bf C #1}{#2}{#3}}
\def\ijmpa#1#2#3{{\it Int.~J.~Mod.~Phys.~}\jref{\bf A #1}{#2}{#3}}
\def\jcp#1#2#3{{\it J.~Comp.~Phys.~}\jref{\bf #1}{#2}{#3}}
\def\jetp#1#2#3{{\it JETP~Lett.~}\jref{\bf #1}{#2}{#3}}
\def\jphysg#1#2#3{{\small\it J.~Phys.~G~}\jref{\bf #1}{#2}{#3}}
\def\jhep#1#2#3{{\small\it JHEP~}\jref{\bf #1}{#2}{#3}}
\def\mpl#1#2#3{{\it Mod.~Phys.~Lett.~}\jref{\bf A #1}{#2}{#3}}
\def\nima#1#2#3{{\it Nucl.~Inst.~Meth.~}\jref{\bf A #1}{#2}{#3}}
\def\npb#1#2#3{{\it Nucl.~Phys.~}\jref{\bf B #1}{#2}{#3}}
\def\nca#1#2#3{{\it Nuovo~Cim.~}\jref{\bf #1A}{#2}{#3}}
\def\plb#1#2#3{{\it Phys.~Lett.~}\jref{\bf B #1}{#2}{#3}}
\def\prc#1#2#3{{\it Phys.~Reports }\jref{\bf #1}{#2}{#3}}
\def\prd#1#2#3{{\it Phys.~Rev.~}\jref{\bf D #1}{#2}{#3}}
\def\pR#1#2#3{{\it Phys.~Rev.~}\jref{\bf #1}{#2}{#3}}
\def\prl#1#2#3{{\it Phys.~Rev.~Lett.~}\jref{\bf #1}{#2}{#3}}
\def\pr#1#2#3{{\it Phys.~Reports }\jref{\bf #1}{#2}{#3}}
\def\ptp#1#2#3{{\it Prog.~Theor.~Phys.~}\jref{\bf #1}{#2}{#3}}
\def\ppnp#1#2#3{{\it Prog.~Part.~Nucl.~Phys.~}\jref{\bf #1}{#2}{#3}}
\def\rmp#1#2#3{{\it Rev.~Mod.~Phys.~}\jref{\bf #1}{#2}{#3}}
\def\sovnp#1#2#3{{\it Sov.~J.~Nucl.~Phys.~}\jref{\bf #1}{#2}{#3}}
\def\sovus#1#2#3{{\it Sov.~Phys.~Usp.~}\jref{\bf #1}{#2}{#3}}
\def\tmf#1#2#3{{\it Teor.~Mat.~Fiz.~}\jref{\bf #1}{#2}{#3}}
\def\tmp#1#2#3{{\it Theor.~Math.~Phys.~}\jref{\bf #1}{#2}{#3}}
\def\yadfiz#1#2#3{{\it Yad.~Fiz.~}\jref{\bf #1}{#2}{#3}}
\def\zpc#1#2#3{{\it Z.~Phys.~}\jref{\bf C #1}{#2}{#3}}
\def\ibid#1#2#3{{ibid.~}\jref{\bf #1}{#2}{#3}}
\def\otherjournal#1#2#3#4{{\it #1}\jref{\bf #2}{#3}{#4}}
\newcommand{\jref}[3]{{\bf #1}, #3 (#2)}
\newcommand{\hepph}[1]{{\tt [hep-ph/#1]}}
\newcommand{\mathph}[1]{{\tt [math-ph/#1]}}
\newcommand{\arxiv}[2]{{\tt arXiv:#1}}
\newcommand{\bibentry}[4]{#1, {\it #2}, #3\ifthenelse{\equal{#4}{}}{}{, }#4.}




\end{document}